\documentstyle{mn}

\input psfig
\catcode`\"=\active\let"=\"
\def\3{\ss }
\def\Stodolkiewicz{Stod\'o\l kiewicz\ }

\def\nbody#1{{\cc Nbody#1}{} }
\def\plusplus{\raise 0.3ex\hbox{${\scriptstyle ++}$}{}}
\def\pmax{p_{\rm max}}
\font\cc cmcsc10

\title[A stochastic Monte Carlo approach ]
{A stochastic Monte Carlo approach to model real star
cluster evolution, III. Direct integrations of three- and four-body interactions.}
\author[M. Giersz and R. Spurzem]
{M.~Giersz$^1$ and R.~Spurzem$^2$ \\
$^1$ Nicolaus Copernicus Astronomical Centre, Polish
  Academy of Sciences, ul. Bartycka 18, 00-716 Warsaw, Poland\\
$^2$ Astronomisches Rechen-Institut,
    M"onchhofstra\3e 12-14, D-69120 Heidelberg, Germany}

\begin{document}  

\maketitle

\begin{abstract}
Spherically symmetric equal mass star clusters containing a large amount of primordial
binaries are studied using a hybrid method, consisting of a gas dynamical
model for single stars and a Monte Carlo treatment for relaxation
of binaries and the setup of close resonant and fly-by encounters of single
stars with binaries and binaries with each other (three- and four-body
encounters). What differs from our previous work is that each encounter is being
integrated using a highly accurate direct few-body integrator which uses regularized 
variables. Hence we can study the systematic evolution of individual binary orbital
parameters (eccentricity, semi-major axis) and differential and total cross sections
for hardening, dissolution or merging of binaries (minimum distance) from
a sampling of several ten thousands of scattering events as they
occur in real cluster evolution including mass segregation of binaries,
gravothermal collapse and reexpansion, binary burning phase and ultimately
gravothermal oscillations. For the first time we are able to present 
empirical cross sections for eccentricity variation of binaries in close
three- and four-body encounters. It is found that a large fraction of three-body
and four-body
encounters results in merging. Eccentricities are generally increased
in strong three- and four-body encounters and there is a characteristic
scaling law $\propto  \exp(4 e_{\rm fin})$ of the differential cross section
for eccentricity changes,
where $e_{\rm fin} $ is the final
eccentricity of the binary, or harder binary for four-body encounters.
Despite of these findings the overall eccentricity
 distribution remains thermal for all binding energies of binaries, which is
 understood from the dominant influence of resonant encounters.
 Previous cross sections obtained by Spitzer and Gao for strong encounters can
 be reproduced, while for weak encounters non-standard processes like formation
 of hierarchical triples occur.

\end{abstract}


\section{Introduction}

 Dynamical modelling of globular clusters and other collisional
 stellar systems (like galactic nuclei, rich open clusters, and
 rich galaxy clusters) still poses a considerable
 challenge for both theory and computational requirements
 (in hardware and software). On the theoretical side
 the validity of certain assumptions
 used in statistical modelling based on the Fokker-Planck
 (henceforth FP) and other approximations is still poorly known.
 Stochastic noise in a discrete $N$-body system
 and the impossibility to directly model realistic particle numbers
 with the presently available hardware, are a considerable
 challenge for the computational side.

A large amount of individual pairwise forces between particles
needs to be explicitly calculated to properly follow
relaxation effects based on the cumulative effects of small
angle gravitative encounters.
Therefore modelling of globular star clusters
over their entire lifetime requires a star-by-star simulation approach,
which is highly accurate in following stellar two- and many-body encounters.
Distant two--body encounters, the backbone of quasi-stationary relaxation
processes, have to be treated properly by the integration method directly,
while close encounters, in
order to avoid truncation errors and downgrading of the overall
performance, are treated in relative and specially regularized
coordinates (Kustaanheimo \& Stiefel 1965, Mikkola \& Aarseth 1996,
1998), their centres of masses
being used in the main integrator. The factual world standard
of such codes has been set by
Aarseth and his codes NBODY$n$-codes ($0\le n \le 7$, see
Aarseth 1985, 1994, 1999a,b, 2003). Recently another integrator {\sc Kira} aimed
at high-precision has been used (McMillan
\& Hut 1996, Portegies Zwart et al. 1998, Takahashi \& Portegies Zwart 1998,
2000). It uses a hierarchical binary tree to handle compact subsystems
instead of regularization.

Unfortunately,
the direct simulation of most dense stellar systems
with star-by-star modelling is not yet possible, although recent
years have seen a significant progress in both hardware and software.
First, parallel (even massively parallel) computing has opened a
route to gain performance at a relatively low cost and little
technological advancement, such as in the case of the
CRAY T3E supercomputers as well as for so-called Beowulf
parallel PC clusters, both being not far from each other
in performance. The fastest supercomputer available in the
German computing centres is a Hitachi SR8000 with 2 Tflop
peak speed. On the other hand side the already successful
GRAPE special purpose computers (Sugimoto et al. 1990, Makino
et al. 1997, Makino \& Taiji 1998, Makino 2002, Makino \& Hut 2003)
have been developed in their present generation (GRAPE-6) and
aim at the 100 Tflops speed. The price to pay for the latter
is small in money (per Tflop) but may be high otherwise
because they are only suitable for special problems and algorithms.
As a general rule one could say that every complexity different
from standard force calculations in an $N$-body system (such as
coupling to gas dynamics, many hard binaries) is a potential
threat for the performance of GRAPE. Also, clever algorithms such
as Ahmad-Cohen neighbour schemes (Ahmad \& Cohen 1973), which
would normally provide much higher efficiency cannot yet be used
in a competitive way on GRAPE. Therefore such codes have
been in the past most efficiently ported on massively parallel
general purpose computers (CRAY T3E and PC clusters); the
prototype code NBODY6++ based on a portable MPI implementation has been
described for the first time by Spurzem (1999) and was used for
astrophysical problems related to the dissolution of globular
clusters (Baumgardt 2001, Baumgardt et al. 2002) and the decay of massive black
hole
binaries in galactic nuclei after a merger (Milosavljevic \& Merritt 2001,
Hemsendorf, Sigurdsson \& Spurzem 2002). It seems that the use of
reconfigurable hardware (Kuberka et al. 1999, Spurzem \& Kugel 1999,
Hamada et al. 2000, Spurzem et al. 2002) helps out to find an
efficient way to use both GRAPE and parallel sophisticated codes.
Another approach to improve the performance of direct $N$-body
codes is the use of systolic and hyper-systolic algorithms
(Lippert et al. 1996, Lippert et al. 1998,
Dorband, Hemsendorf, \& Merritt 2002 Makino 2003).

Despite such progress in hardware and software even
the largest useful direct $N$-body models for both globular
cluster and galactic nuclei
evolution have still not yet reached the realistic
particle numbers ($N \sim 5\times 10^5$ for globular clusters,
$N \sim 10^6$ to $10^9$ for galactic nuclei). However,
recent work by Baumgardt, Hut \& Heggie (2002) and Baumgardt \& Makino
(2002) has pushed the limits of present direct modelling for the
first time to some $10^5$ using both NBODY6++ and
the GRAPE-6 special purpose hardware.
There is a
notable exception of a direct one-million body problem with
a central binary black hole tackled by Dorband, Hemsendorf \ Merritt
(2001). Their code, however, being very innovative for the large
$N$ force calculation, still lacks the fine ingredients of treating
close encounters between stars and black hole particles from
the standard $N$-body codes.

Bridging the gap between direct models and the most interesting particle
numbers in real systems is far from straightforward, neither by scaling
(Aarseth \& Heggie 1998, Baumgardt 2001), nor by theory.
There are two main classes of theory: (i) FP models,
which are based on the direct numerical solution of the orbit-averaged
FP equation
(Cohn 1980, Cohn, Hut \& Wise 1989, Murphy, Cohn \& Hut 1990,
Murphy, Cohn \& Durisen 1991),
and (ii) isotropic (Lynden-Bell \& Eggleton 1980,
Heggie 1984, Bettwieser \& Sugimoto 1984) and
anisotropic gaseous models
(Louis \& Spurzem 1991, Spurzem 1994, 1996),
which can be thought of as a set of
moment equations of the FP equation.

 Detailed comparisons for equal mass isolated star
 clusters have been performed (Giersz \&
 Heggie 1994a,b, Giersz \& Spurzem 1994,
 Spurzem \& Aarseth 1996, Spurzem 1996)
 with direct $N$-body simulations using standard $N$-body codes
 (\nbody{5}, Aarseth 1985, Spurzem \& Aarseth 1996; \nbody{2},
 Makino \& Aarseth 1992; \nbody{4}, Makino 1996; \nbody{6\plusplus},
 Spurzem 1999). There have yet been very
 few attempts to extend the quantitative comparisons to
 more realistic star clusters containing different mass bins or
 even a continuous mass spectrum (Spurzem \& Takahashi 1995,
 Giersz \& Heggie 1996).

 On the side of the FP models there have been two major
 recent developments. Takahashi (1995, 1996, 1997) has published
 new FP models for spherically symmetric star clusters,
 based on the numerical solution of the orbit-averaged 2D FP
 equation (solving the FP
 equation for the distribution $f=f(E,J^2)$ as a function of
 energy and angular momentum, on an $(E,J^2)$-mesh).
 Drukier et al. (1999) have published
 results from another 2D FP code based on the original
 Cohn (1979) code. In such 2D FP models
 anisotropy, i.e. the possible difference between radial
 and tangential velocity dispersions in spherical clusters, is taken
 into account. Although the late, self-similar stages of core collapse
 are not affected very much by anisotropy (Louis \& Spurzem 1991),
 intermediate and outer zones of globular clusters, say outside roughly
 the Lagrangian radius containing 30\% of the total mass, do exhibit
 fair amounts of anisotropy, in theoretical model simulations as well
 as according to parametrized model fits (Lupton, Gunn \& Griffin 1987).
 In contrast to the anisotropic gaseous models the 2D FP
 models contain less inherent model approximations; they do not
 assume a certain form of the heat conductivity and closure relations
 between the third order moments as in the case of the anisotropic
 gaseous model. Furthermore, the latter contains a numerical
 constant $\lambda $ (Spurzem 1996), which is of order unity,
 but its numerical value has to be determined from comparisons with
 proper FP or $N$-body models.

 Secondly, another 2D FP model has been worked out recently
 for the case of axisymmetric rotating star clusters (Einsel \& Spurzem
 1999, Kim et al. 2002). Here, the distribution function is assumed to be a function
 of energy $E$ and the $z$-component of angular momentum $J_z$ only;
 a possible dependence of the distribution function on a third integral
 is neglected. As in the spherically symmetric case the neglection
 of an integral of motion is equivalent to the assumption
 of isotropy, here between the velocity dispersions in the meridional
 plane ($r$ and $z$ directions); anisotropy between velocity
dispersion in the meridional plane
 and that in the equatorial plane ($\phi$-direction), however, is
included.

 With the advent of a wealth of detailed data on globular clusters,
 such as luminosity functions and derived mass functions,
 color-magnitude diagrams, and population and kinematical analysis,
 obtained by e.g. the Hubble Space Telescope, for extragalactic
 as well as Milky Way clusters (cf. e.g. Piotto \& Zocalli 1999,
 Rubenstein \& Bailyn 1997, Ibata et al. 1999, Piotto et al. 1999,
 Grillmair et al. 1999, Shara et al. 1998, to mention only the few
 most recently appeared papers), easily reproducible reliable
 modelling becomes more important than before. For that purpose
 a few more ingredients are urgently required in the models in
 addition to anisotropy and rotation: a
 mass spectrum, a tidal field, and the influence of stellar evolution on
 the dynamical evolution of the cluster.

 In principle it appears easy to include all these in a direct $N$-body
 simulation, but it turned out that considerable effort is needed to
 improve the required physical knowledge and numerical codes
 for that purpose (Hut \& Makino 1999, Makino et al. 1997, 
 Aarseth 1999). A notable effort to focus the experience of
 different groups here is the recent MODEST initiative (Hut et al. 2002,
 Sills et al. 2003 and webpage at {\tt http://www.manybody.org/modest.html}).
 Since the life span of globular clusters extends over
 many tens of thousands of crossing times, taken at the half-mass
 radius,
 their simulation requires a highly accurate
 direct $N$-body code.
 Despite the enormous advances in hardware and software
 the modelling of say 100.000 particles is still presently a very
 challenging task and it is impossible to do large parameter surveys
 in this regime, even with NBODY6++ or GRAPE-6. One has
 to rely on lower particle numbers and prescriptions to scale the
 results to larger $N$ (Aarseth \& Heggie 1998, Baumgardt 2001). So we still need
 the fast but approximate theoretical models.

  But there is an elegant alternative way to generate models of
 star clusters, which can correctly reproduce the stochastic features
 of real star clusters, but without really integrating
 all orbits directly as in an $N$-body simulation. These so-called Monte Carlo
 models were first presented by H\'enon (1971, 1975), Spitzer (1975) and later
 improved by \Stodolkiewicz (1982, 1985, 1986) and in further
 work by Giersz (1996, 1998, 2001). The basic idea is
 to have pseudo-particles, whose
 orbital parameters are given in a smooth, self-consistent potential.
 However, their orbital motion is not explicitly followed; to
 model interactions with other particles like two-body relaxation
 by distant encounters or strong interactions between binaries and
 field stars, a position of the particle in its orbit and further
 free parameters of the individual encounter are picked from
 an appropriate distribution by using
 random numbers.

 There are also Monte-Carlo models
 by Joshi, Rasio \& Portegies Zwart (2000),
 Watters, Joshi \& Rasio (2000), Joshi, Nave \& Rasio (2001), 
 Fregeau et al.(2003) for globular
 clusters and Freitag (2000), Freitag \& Benz (2001, 2002) for globular
 clusters and galactic
 nuclei. They rely on the FP approximation and (hitherto) spherical
 symmetry, but their data structure is very similar to a $N$-body model.

 A hybrid method using a gas dynamical model for
 the single stellar component, but a Monte Carlo technique for the
 binaries has been proposed in Paper I (Spurzem \& Giersz 1996) and
 it was shown in Paper II (Giersz \& Spurzem 2000) that such models are in
 good agreement with $N$-body and FP results where it should be
 expected, and that they provide a huge amount of interesting, detailed
 informations about binary evolution in a star cluster (globular cluster)
 with a realistically large number of primordial binaries and single stars
 (30k binaries in cluster with 270k single stars). Another attempt to
 combine a direct FP model with Monte Carlo to a hybrid scheme
 has been reported but not continued further yet (Huang \& McMillan 2001).
 In this paper we
 present results obtained from a further improvement of our code which
 allows us to study in a much more realistic way the evolution of binaries
 in star clusters by using direct three- and four-body integration
 in regularized variables to follow the close encounters.

 It cannot be excluded that star formation in clusters predominantly
 produces tight binaries rather than single stars, and this is
 consistent with present day binary fractions in the galactic field
 \cite{1995MNRAS.277.1491K}.
 Globular cluster observations reveal present day binary
 fractions of about 15\% or more, depending on the position in the
 cluster (Rubenstein \& Baily 1997). Regarding the large dynamical age
 of (at least in the Milky Way) globular clusters we must assume that
 stars in them also form with a high fraction of binaries. As was reported in
 more detail in Paper II the dynamical modelling of large star clusters
 with many primordial (=initial) binaries, many of them hard (i.e. with
 binding energies higher than the average thermal energy of stars in
 the cluster, often denoted with $kT$), is an even harder physical
 and computational task as the standard $N$-body models. From the viewpoint
 of realism and the aim to finally obtain models comparable to real
 clusters the main improvement here, in Paper III, is that for the first
 time we have detailed information about all orbital elements of the
 binaries, in particular their eccentricities.

 While semi-major axis or binding energies of binaries have been subject
 to intensive study since several decades, much less is known about
 the systematic evolution of eccentricities of binaries in cluster
 environments. In a pioneering study Hills (1975a,b) simulated about
 10.000 encounters between binaries and single stars (equal masses)
 and discussed how
 binaries would interact with a star cluster environment. He found
 that the eccentricity of initially circular
 binaries after a strong encounters would
 approach the thermal mean of $\langle e \rangle = 2/3 $ for slow
 encounters, and that moderately fast encounters provide even larger
 eccentricities. With a few thousand direct integrations of three-body scatterings
 Hut \& Paczy\'nski (1984) determined first cross sections for eccentricity change
 of binaries with circular orbits for wide encounters.
 Sigurdsson \& Phinney (1993) performed stochastic
 binary encounters in a static cluster background (again, only encounters
 with circular binaries are considered) with much better statistics and
 unequal masses. Their results are consistent with the earlier work
 of Hills (1975a,b) regarding the eccentricities, that distant fly-by
 encounters generate little changes in eccentricities, while close and
 exchange encounters tend to generate on average a high thermal eccentricity
 value of $\langle e \rangle = 2/3 $. McMillan et al. (1991) performed small $N$-body
 simulations (100 binaries in a 1100 star model) and for the first time approximately
 determined differential cross sections for three- and four-body interactions in
 an evolving star cluster. They also, followed evolution of eccentricity of interacting
 binaries. Their results are consistent with Heggie's (1975) analytical predictions
 for differential cross sections and with thermal eccentricity distributions.

 To assess, however, the steady state distribution of eccentricities in
 a globular cluster we need to know the statistical variation of eccentricities
 in encounters with binaries having initially all possible eccentricities
 (in fly-by and exchange), and we need to know how the evolving cluster
 background interacts with the binaries' eccentricity distribution.
 Such information is crucial to judge about the frequency and dynamical
 importance of some very important physical processes in clusters, such
 as the rate of mergers (producing blue stragglers), the rate of X-ray
 binaries and other exotic objects (binary pulsars, their decay rate due
 to gravitational radiation, see e.g. Rasio \& Heggie 1995),
 the formation and evolution of products of
 close tidal interactions. All these critically depend on the eccentricities
 of the binaries involved in them. Unfortunately, to the knowledge of the
 authors, there is nearly no systematic study of the evolution of initially
 non-circular binaries with the remarkable exception of Heggie (1975),
 McMillan et al. (1991) and Heggie \& Rasio (1996) and some unpublished work
 quoted in the latter paper. Heggie \& Rasio (1996) give cross sections 
 of eccentricity changes using
 an analytic approximation for very distant encounters (tidal and slow) --
 what is most interesting here for us is that they find for initially non-zero
 eccentricity an equal cross section for positive and negative changes of
 eccentricities. So, the earlier picture that due to close encounters binaries
 tend to obtain a certain average eccentricity is complemented by wide
 encounters, which seem to generate a picture of a diffusion process where
 negative and positive changes of eccentricity are alike.

 It is one of the aims of this paper to provide the missing link between
 the two pictures and discuss the complete binary evolution, including
 their eccentricities, covering all initial values, fly-bys and exchanges,
 and incorporate this in an evolving cluster background. The main challenge
 here, besides the development of our stochastic Monte Carlo model in
 general is to disentangle the wealth of data produced by a model back into
 meaningful cross sections and expectations for eccentricity evolution
 (for example).

 In the following section we describe in some detail how we
 initialize and perform the direct integrations for binary-single
 and binary-binary encounters and how the outcome of the encounters
 is used as an input for our stochastic Monte Carlo of the evolving
 cluster. Sect. 3 describes some results, and Sect. 4 gives
 discussion and conclusion.

\section{Binaries and direct three- and four-body integrations}

 In Paper II we used statistical cross sections derived from
 theory or from simplified numerical studies for the changes in binding
 energy of a hard binary in a three- or four-body encounter (Spitzer 1987,
 Gao et al. 1991). Here our aim is to relax this rather severe
 limitation of our models, for two main reasons. One is that such cross
 sections are poorly known (if at all) for the most interesting case of
 systems with unequal masses (spectrum of stellar masses). To prepare
 the simulation of large stellar systems, including a large number
 of primordial binaries and a mass spectrum, in our
 stochastic Monte Carlo model in the next Paper IV we have to
 implement and test the relevant procedures first in the equal mass case,
 which is the subject of this paper. Even in the equal mass case it is
 unknown how the eccentricity of the binaries changes during close
 three-body and four-body encounters, and how the differential and total
 cross sections depend on the initial eccentricity of the binaries which
 react with each other. Eccentricities of binaries play an important
 role for the minimum distance of two stars during an encounter, which will
 be an important tool to study the merging of stars in clusters due to
 close encounters.

 Our present approach is as follows: different from Paper II we use a
 completely unbiased prescription to determine whether a binary-binary encounter
 is due in the sense that the probability does not assume any prescribed
 form of the total cross section. Rather, we obtain a maximum impact
 parameter and use encounter timescales based on it (see details below).
 For three-body encounters we still use
 a total cross section as in Paper II to determine whether an encounter
 takes place, since we check that our results
 correctly reproduce the known cross sections of Spitzer (1987),
 so it is less critical here to adopt a more general encounter probability.
 However, this works only for the equal mass
 case. For the test runs (discussed in more details at the end of section 2.1.1)
 we used the unbiased determinations of three- and four-body encounter
 probabilities. In future simulations for unequal mass systems we will only
 use the unbiased prescription.
 If an individual close encounter of the three or four bodies is due,
 we follow the trajectory
 using regularized coordinates and precise $N$-body integration. After some time the
 reaction products separate again and we measure the outcome (single stars,
 binaries) and the internal and external energies, plus eccentricities of
 the binaries.  Also it is possible to make some statements about the
 relative importance of the formation of hierarchical stable triples.

\subsection{Direct integration of three- and four-body interactions}

\subsubsection{Decision whether encounter takes place}

 At each time step of the gas model it is checked by a random number
 check whether a close encounter is due between a single star and a binary,
 or between two binaries. If a single star is needed, its parameters in
 the cluster frame are drawn from a Schwarzschild-Boltzmann velocity
 distribution, whose parameters (density, radial and tangential velocity
 dispersion) at the location of the binary are given by the gaseous model
 variables. For binaries which are subject to a possible encounter with 
 other binaries the nearest binary neighbour
 is chosen as an interaction partner (Stodolkiewicz's 1982,1986).

 Whether a binary actually suffers from a close three- or four-body encounter
 is determined randomly for each timestep and binary. For three-body encounters
 a probability based on the total cross section of Spitzer (1987) and the
 local density of single stars is used as in Paper II, see above. For four-body
 encounters a new unbiased procedure is used, based on the determination of
 a reasonably large maximum impact parameter $\pmax$ to cover all relevant
 encounters.
 For the determination of $\pmax$ we follow the prescriptions
 outlined in papers of Hut \& Bahcall (1983),
 Mikkola (1983, 1984a,b) and
 Bacon, Sigurdsson \& Davies (1996), with a small modification described below.
 It means we use following prescription

 \begin{equation}
 \pmax = a (D + C/V)
 \end{equation}
 where $V=v/v_c$ is a scaled relative velocity at infinity (using a
 critical velocity $v_c$, see below),
 $a$ and $e$ are the semi-major axis and eccentricity of the
 binary involved in a three-body encounter, or in the case of a four-body encounter the
 values taken from the weaker binary. For
 three-body interactions we use $D=0.6 (1+e)$, $C=4$,
 a choice which represents an empirical optimum, to include
 all relevant impact parameters for changes of the binary parameters, but to cut off too
 many very weak interactions which would not significantly change the internal binary
 parameters (note that two-body relaxation due to the cumulative effect of very
 many distant encounters between single stars and centres of masses of binaries
 is taken properly into account, see Paper II). It is clear, that interactions with
 very high relative velocity only matter for the binary parameters if there is a
 relatively small impact parameter, but on the other hand slow encounters involving wide
 and/or eccentric binaries (large apocentre separation of the binary!) need much more
 caution to catch all relevant encounters, i.e. larger $\pmax$ is necessary. These competing
 effects are accommodated in the terms containing $C$ and $D$, respectively, above.

 The critical velocity in the case of the three-body problem is given by
 \begin{eqnarray}
 v_c^2 &=& { m_1+m_2+m_3 \over m_3 (m_1+m_2) } \cdot\Bigl( {G m_1 m_2\over a} \Bigr)
       ={ 2G \over \mu_3 } E_{b}\cr
 \mu_3 &=& {m_3 (m_1+m_2) \over (m_1+m_2+m_3)}  \ .
 \end{eqnarray}
 Here $m_1$, $m_2$ are the masses of the two members of the binary, $m_3$ is the mass
 of the incoming perturber, single star in case of three-body encounters,
 $E_{b}$ is the binding energy of the binary, and
 $\mu_3$ is a reduced mass.

 In the case of four-body encounters we use $D=0.6 (1+e_1)$, $C=5$, following
 Bacon, Sigurdsson \& Davies (1996), with the four-body critical
 velocity

 \begin{eqnarray}
 v_c^2 &=&  {m_1+m_2+m_3+m_4 \over (m_1+m_2)(m_3+m_4)}  \cdot
   \Bigl( { G m_1 m_2 \over a_1} + {G m_3 m_4 \over a_2 }\Bigr)  \cr
      &=&   { 2G \over \mu_4 } ( E_{b1} + E_{b2} ) \cr
  \mu_4 &=& {(m_1+m_2)(m_3+m_4) \over (m_1+m_2+m_3+m_4)} \ .
 \end{eqnarray}
 Now $m_1$, $m_2$, $a_1$, $e_1$, $E_{b1}$ denote parameters of the weaker binary
 in a four-body binary-binary encounter, $m_3$, $m_4$, $a_2$, $e_2$, $E_{b2}$
 and those of the harder binary, $\mu_4$ is a reduced mass.

 From $p_{\rm max}$ we determine an encounter rate $\dot{N}$ using the simple
 prescription $\dot{N}=nAv_{\rm rel}$, where $n$, $A=\pi p_{\rm max}^2$, $v_{\rm rel}$ are
 the local binary particle density, the geometrical cross section of
 $p_{\rm max}$, and the actual relative velocity of the two individual binaries
 chosen for a close encounter. It is crucial that the binary
 density is well-defined, smooth and steady, otherwise spurious fluctuations
 will cause convergence problems in the gaseous model equations and other
 problems. We finally determine the binary density by smoothing out each
 binary over its accessible coordinate space (weighted with the orbital
 factor $dr/v_r$), and summing all binaries' contributions in each shell
 to obtain the local $n$.
 To determine, in a given volume $\Delta V $ (radial shell of
 our gaseous model) and time step of the simulation $\Delta t $,
 whether an encounter takes place we
 compare $\dot{N} \Delta t$ with a random number.

 When setting up our simulation and testing the code it turned out that one
 has to make some small modifications of the above standard procedure for
 the purpose of computing speed. With $p_{\rm max}$ as defined above we find
 that our program integrates in a typical run, ranging over dozens of half-mass
 relaxation times and thousands of binaries, (initially tens of thousands)
 several hundred thousands of weak four-body encounters, which lead to very small changes in
 the binary parameters. The computing time is dominated by the quadruple integration
 in an unpracticable and undesired way. Therefore we limited, in the
 case of binary-binary encounters, the maximum impact parameter to
 $p_{\rm max} = 10 a_1$ (semi-major axis of weaker binary). In several
 control runs, we checked what is the influence of such a limitation
 on the global dynamical evolution. Unfortunately there is no simple
 conclusion from these experiments. While many features of global evolution
 (order of first time of collapse, minimum and maximum density during later
 gravothermal oscillations) are similar, we get some different results in
 the detailed binary evolution. Generally fewer hard binaries develop, there
 is less activity caused by them, e.g. the number of escapers (single and
 binary) drops by some 10 \% if the full $p_{\rm max}$ is used, as compared
 to our standard run, using the limitation of $p_{\rm max}$. However,
 using large $p_{\rm max}$ is not necessarily the correct way to solve the
 problem, because at some point weak encounters are identical to standard
 two-body relaxation between two binaries and this is treated otherwise -
 by our standard Monte Carlo procedure. In view of these difficulties we
 decided to present here our results from a standard run using the
 $p_{\rm max}$ limitation as described above, while making the reader
 here aware of the underlying possible problems. The solution to the ''$p_{\rm max}$''
 problem will be a better termination criterion for three- and four-body
 encounters, which will limit the integration time for rare but time consuming
 cases. Unfortunately, we have not yet found any reliable termination criterion.
 It is worth to note, that
 simulations up to the time equal to about a few hundred initial relaxation times
 take about two weeks and about two months with limited and unlimited
 $p_{\rm max}$, respectively.

\subsubsection{Setting up an encounter}

 After we have identified three or four bodies to interact, given their
 masses and velocities at infinity in their centre of mass frame, there is
 some sequence of steps until the actual three- or four-body integration should start.

 Firstly, using a random sampling of impact parameter $p$ assuming a constant distribution
 in $dp^2 = 2\pi p dp $, with $0 < p < p_{\rm max}$ we get the actual impact parameter
 to be used for this encounter. Secondly,
 the detailed integration of each encounter starts at some finite separation
 $r_x$ of the reaction partners, where the perturbation of the weaker binary
 becomes relevant. We determine $r_x$ by the condition that
 the weaker binary experiences a force perturbation $\gamma > \gamma_{\rm crit}$ with
 $\gamma_{\rm crit} = 10^{-3}$, and allowing for some generous fudge
 factors to make sure that under any orbital phase we still have $\gamma < \gamma_{\rm crit}$
 initially. In case of three-body encounters we use such a criterion analogously for
 the perturbation of the binary by a single star. To be specific, we take

  \begin{equation}
  r_x = A \cdot a (1+e) \Biggl( 1 + \Bigl( {8 m_x \over \gamma_{\rm crit} (m_1+m_2)}\Bigr)^{1/3}\Biggr)
  \end{equation}
 where $a$, $e$ are again semi-major axis and eccentricity of the binary (three-body)
 or of the
 weaker binary (four-body),
 and $A=10$ (three-body) or $A=3.5$ (four-body) are the before mentioned fudge factors.
 In case of four-body encounters $m_x = m_3 + m_4$, whereas for a three-body encounter
 $m_x = m_3$.
 Note, that we present such procedures here already using individual masses
 $m_i$ for all stars taking part in the encounters, in order to prepare the ground for
 subsequent work for multi-mass systems. All results presented in this paper are
 obtained for systems where $m_i = $ const. $ = 1/N$ in normalized variables.

 \begin{figure}
 \psfig{figure=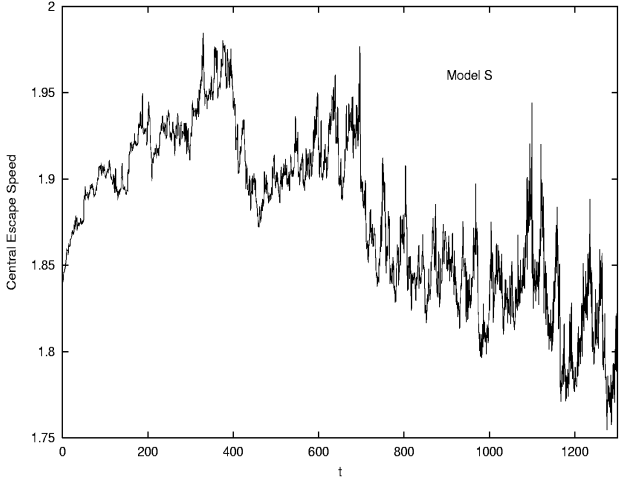,height=6.5cm,width=8.5cm,angle=-90}
 \caption {Model S, central escape speed as a function of time (in $N$-body units).}
 \label{f1}
 \end{figure}

 After using energy and momentum conservation to transform the encounter from
 infinity to $r_x$ we start the actual three- or four-body integration at
 separation $r_x$; first all variables are transformed into a frame in which
 the centre of mass of the three-body (four-body) system is at rest. Then
 regularized coordinates are obtained using Aarseth \& Zare (1974) for three-body
 and Mikkola (1983) for four-body regularization. We use the integration
 packages TRIPLE and QUAD for unperturbed three- and four-body integration,
 which were in their original version kindly supplied by S.J.~Aarseth with his
 NBODY6 program package. In both packages the minimum distance between all
 interacting particles is correctly
 tracked, even if the steps in regularized time do not exactly match with
 the pericentre of the motion of two bodies. Secondly termination is checked
 primarily against a maximum separation of the outermost body; for this
 parameter we choose here $1.2 r_x$, where $r_x$ is the initial separation
 at the setup of the encounter as defined above. In some rare cases where
 the state of the remaining three bodies is not well defined we continue the
 integration until the next largest distance also reaches $1.2r_x$ or a
 well defined end-state is reached (bound triple). If a bound quadruple is
 formed (three-body encounters: bound triple) we also terminate the integration.

\begin{figure}
 \psfig{figure=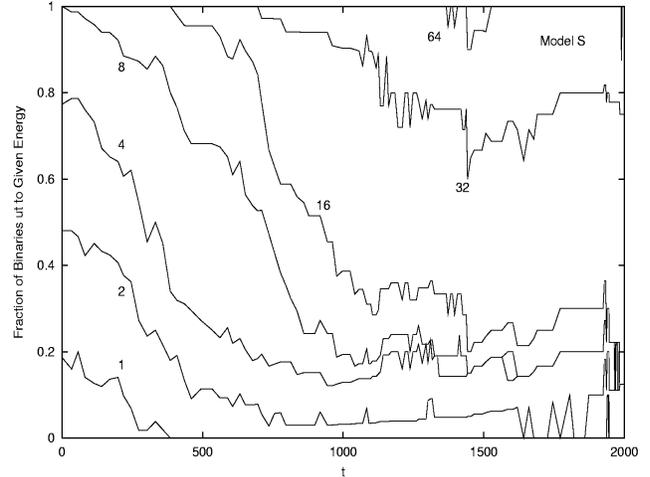,height=6.5cm,width=8.5cm,angle=-90}
 \caption{Model S, distribution of binding energies as
 a function of time (in $N$-body units) for all bound binaries. The unit of energy is
 $2kT$, i.e. one and a half times the initial
 mean kinetic energy of the single stars. Each line gives the fraction of these
 binaries with binding energy smaller than the stated value.}
 \label{f2}
 \end{figure}

 \begin{figure}
 \psfig{figure=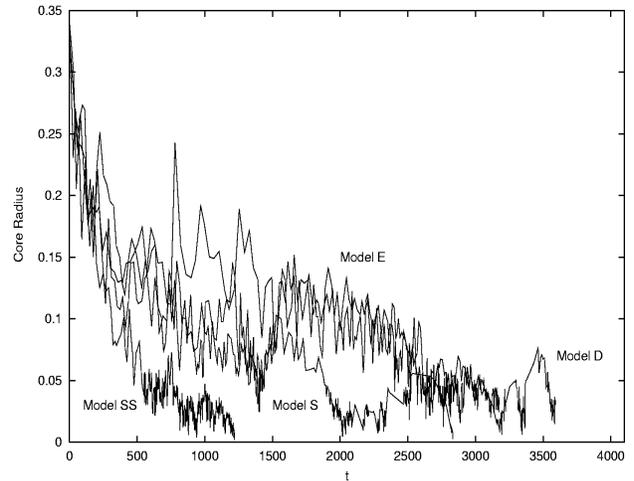,height=6.5cm,width=8.5cm,angle=-90}
 \caption {Models S, D, E and SS; core radius as a function of time (in $N$-body units).}
 \label{f3}
 \end{figure}

 \begin{figure}
 \psfig{figure=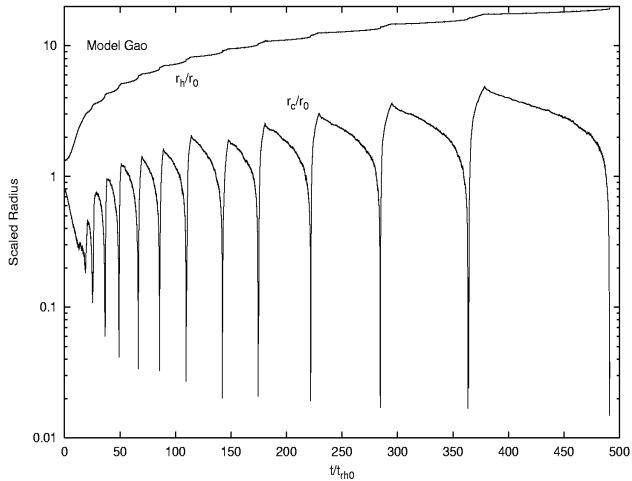,height=6.5cm,width=8.5cm,angle=-90}
 \caption{Evolution of the core and half-mass radii (scaled by
 the scale length of the Plummer model) as a function of time (scaled by the
 initial half-mass relaxation time).}
 \label{f4}
 \end{figure}

 If TRIPLE and QUAD terminate with the above distance criterion we check
 the resulting new configuration after the encounter. We adopt the following
 channels for the outcome of the reaction in the case of three-body encounters:

\begin{itemize}
\item {\it 1 hardening:} the remaining two stars are gravitationally bound with well
 defined semi-major axis and eccentricity. The third star carries away surplus
 kinetic energy; this is the typical case of superelastic scattering which
 heats the stellar system.
\item {\it 2 softening:} the binary loses binding energy at the expense of the
 single star. This and the following two cases are rare, and occur predominantly
 for weak binaries.
\item {\it 3 three bound:} a bound three-body subsystem is formed. For practical
 reasons we define a threshold of $0.1kT$; only if the three-body system in
 total is bound with more than that value, it is considered bound.
\item {\it 4 dissolution:} the kinetic energy of the incoming star was large enough
 to disrupt the binary, at the end there are three single stars absorbed into
 the single star component.
\end{itemize}

 While cases 1,2,4 can be treated appropriately in all changes of their
 physical quantities (mass, internal and external energies of all
 reaction partners and products are properly balanced from and to single
 and binary components of our system), case 3 cannot yet be correctly treated.
 In future work we will also keep track of bound triples, which then would
 be subject to further interactions, requiring eventually direct $N$-body
 integration with more than four bodies, e.g. by using the CHAIN method
 (Mikkola \& Aarseth 1996, 1998). At present we just terminate the triple
 and convert it back into a binary (obtained from its two most bound stars)
 and a single star. The binary binding energy is reduced by the corresponding
 amount to lift the escaping star to zero energy. While this procedure is
 artificial for this particular triple it ensures that there is no
 energetic error generated. The fraction of such events is very small.

 For four-body encounters we employ these criteria to judge about their
 outcome and reaction products:

\begin{itemize}
\item {\it 1 stable:} two binaries are well defined and bound after the
 encounter. They part from each other after the encounter with changed orbital
 parameters. The most probable case is that the harder binary becomes harder,
 and in about one third of the encounters an exchange takes place.
\item {\it 2 dissociation:} one binary is dissolved, mostly it is the weaker
 binary. It is not excluded, however, especially if both binaries have similar
 binding energies, that only the harder binary is dissolved.
\item {\it 3 hierarchical:} one star escapes, and a hierarchical triple remains;
 here we check whether it fulfills the stability criterion of Mardling \& Aarseth (2001).
 If not, the integration is continued for some time, until the third star
 has also escaped from a remaining binary. In very rare cases we have to terminate
 an encounter,
 because after long integration time neither a hierarchical and stable triple
 nor a binary will reach the escape criterion. It is interesting to note here that
 stable three-body orbits exist, which are not hierarchical (Heggie 2000); in this
 work, however, we did not check in detail the outcome of such situations, which
 should be done in the future in a framework of treatment of triples in our
 hybrid scheme.
\item {\it 4 four bound:} all four stars form a bound subsystem.
\end{itemize}

 In the framework of our model we treat properly cases 1 and 2, as the most
 frequent ones. Cases 3 and 4 are usually decomposed into a binary and single
 stars forcibly, with an analogous procedure as described above for three-body
 encounters in order to guarantee correct energetics. In the future
 we will be able to deal with hierarchical triples in our
 stochastic Monte Carlo model.

 \begin{figure}
 \psfig{figure=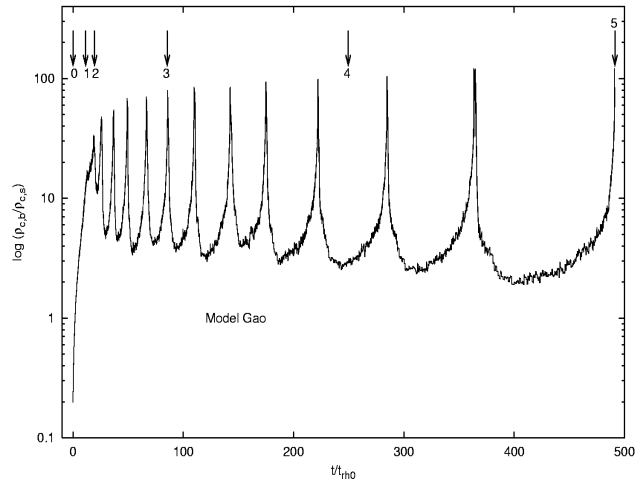,height=6.5cm,width=8.5cm,angle=-90}
 \caption{Ratio of central densities in binaries and singles as a
 function of time (scaled by the initial half-mass relaxation time).
 Arrows indicate the times for snapshot plots of 2D
 distributions of binaries bound to the system (see Fig.~\ref{f6}).}
 \label{f5}
 \end{figure}

 Each full star cluster run stores a data bank of all relevant information for
 each encounter that occurred during the run. A typical amount of data comprises
 some 20k four-body interactions and 40k three-body interactions. To test
 our limitation of $p_{\rm max}=10 a_1$ (see last paragraph of Sect. 2.1.1) we
 did two models where we allowed for {\it all} three- and four-body encounters to be
 directly integrated within $p_{\rm max}$ as deffined in eq. (1), no matter how weak they are.
 Each run comprised of about 70k and 500k integrated three- and four-body encounters,
 without significant
 differences to the standard one (except a better coverage of weak fly-bys, which
 could be seen by an increase of the cross sections for very small changes).

 Since we
 are interested in the statistics of binary evolution in the framework of real
 cluster evolution, we do not use the strategy to sample individual three-
 or four-body
 encounters in artificial settings. Rather, we simulate how real binaries suffer
 from encounters in an evolving cluster, but process the data of all binary scatterings
 afterwards in a very similar way as compared to Hut \& Bahcall (1983) and
 Bacon, Sigurdsson \& Davies (1996).
 By postprocessing of the data bank we can also
 look at certain cross sections, e.g. for having a minimum distance smaller than
 a solar radius in order to check for possible mergers. Mergers are not processed
 in our model, we continue to treat all stars as point-mass objects.
 Looking at the fraction of events which would lead to mergers gives us nevertheless
 some good indication of the expected rates of them.

 For the determination of total and differential cross sections we count all
 events, and sort them into the different categories, using the analogous normalizations
 as in the cited papers.

\section{Results}

\subsection{Heggie's Runs, Testing}

 First we repeat a number of models simulated by Heggie \& Aarseth (1992),
 which we call Heggie's runs. They published results of
 full direct $N$-body simulations of 2500 particles with initial binary
 numbers of 75 (models S, SS) or 150 (models D,E), and initial binding
 energy ranging from 2 to 20 $kT$ (models S,D), 2 to 200 $kT$ (model E), and
 2 to 2000 $kT$ (model SS). These models have been carefully examined again
 in our Paper II using analytical cross sections of binary encounters,
 and again we have, with our more detailed binary modelling, repeated the
 series now. Generally all results agree fairly well, with Paper II as well as
 with the original Heggie \& Aarseth (1992) paper. While we do not want to show all results here
 in detail, a few interesting improvements could be noticed.

 Fig.~\ref{f1} shows the central escape speed as a function of time in model S.
 Its maximum value is higher than in Paper II, and the time of the maximum is
 later. Both features make it more similar to the $N$-body results. Another
 difference can be seen in Fig.~\ref{f2} for model S, where there is a much smaller number
 of hard binaries (line representing 64 $kT$) in late times
 as compared to Paper II. This is due to the much more efficient disruption
 of binaries in the initial phase by four-body encounters. Fig.~\ref{f3},
 showing the core radii of the whole system also shows a much smaller
 core radius now for model D (150 binaries) than before in Paper II, again
 because of less heating provided by the binaries. All features discussed in this
 paragraph make the results of our simulations more similar to Heggie \& Aarseth (1992).

\begin{figure*}
\vskip 19truecm
\includegraphics{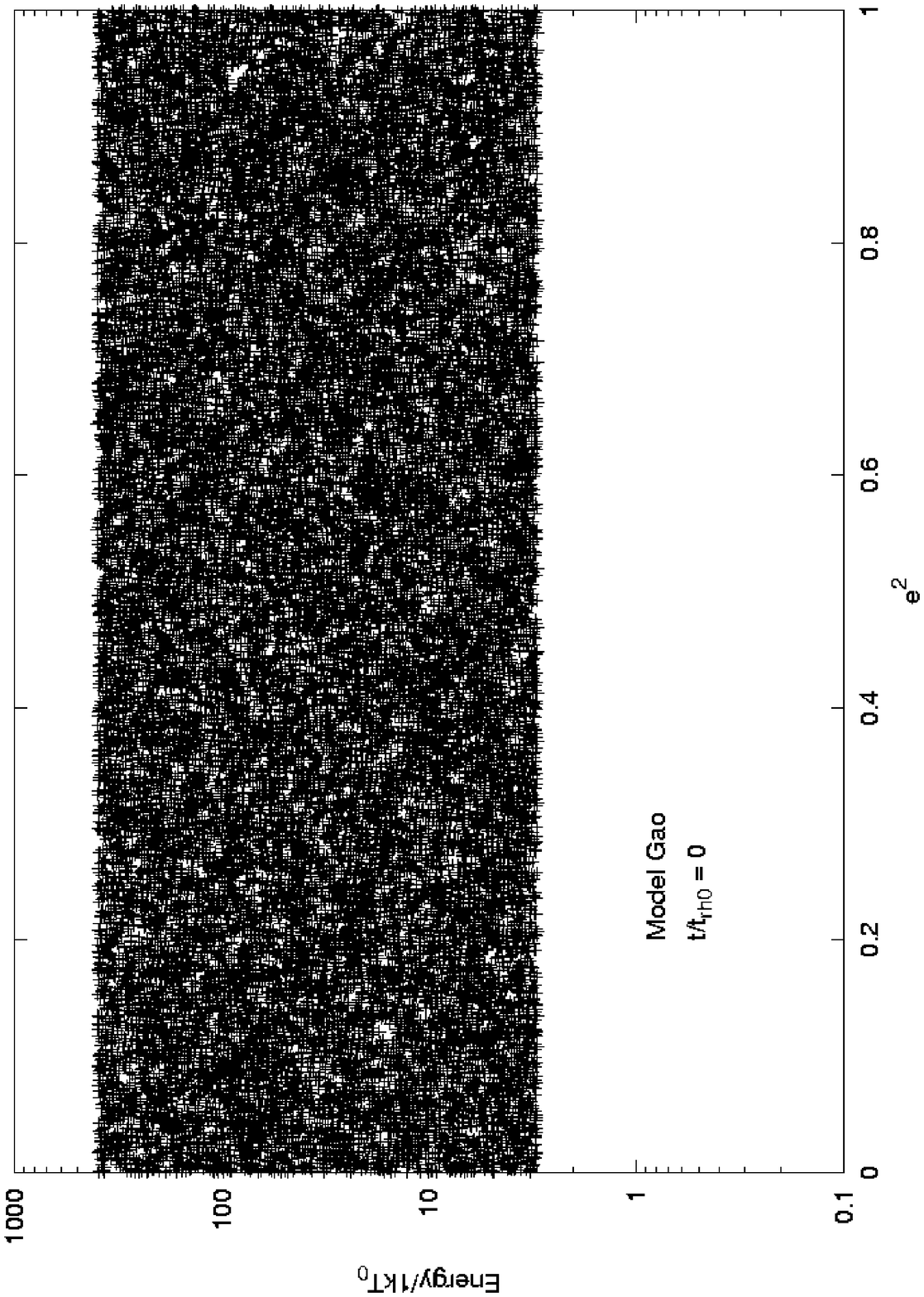}
\includegraphics{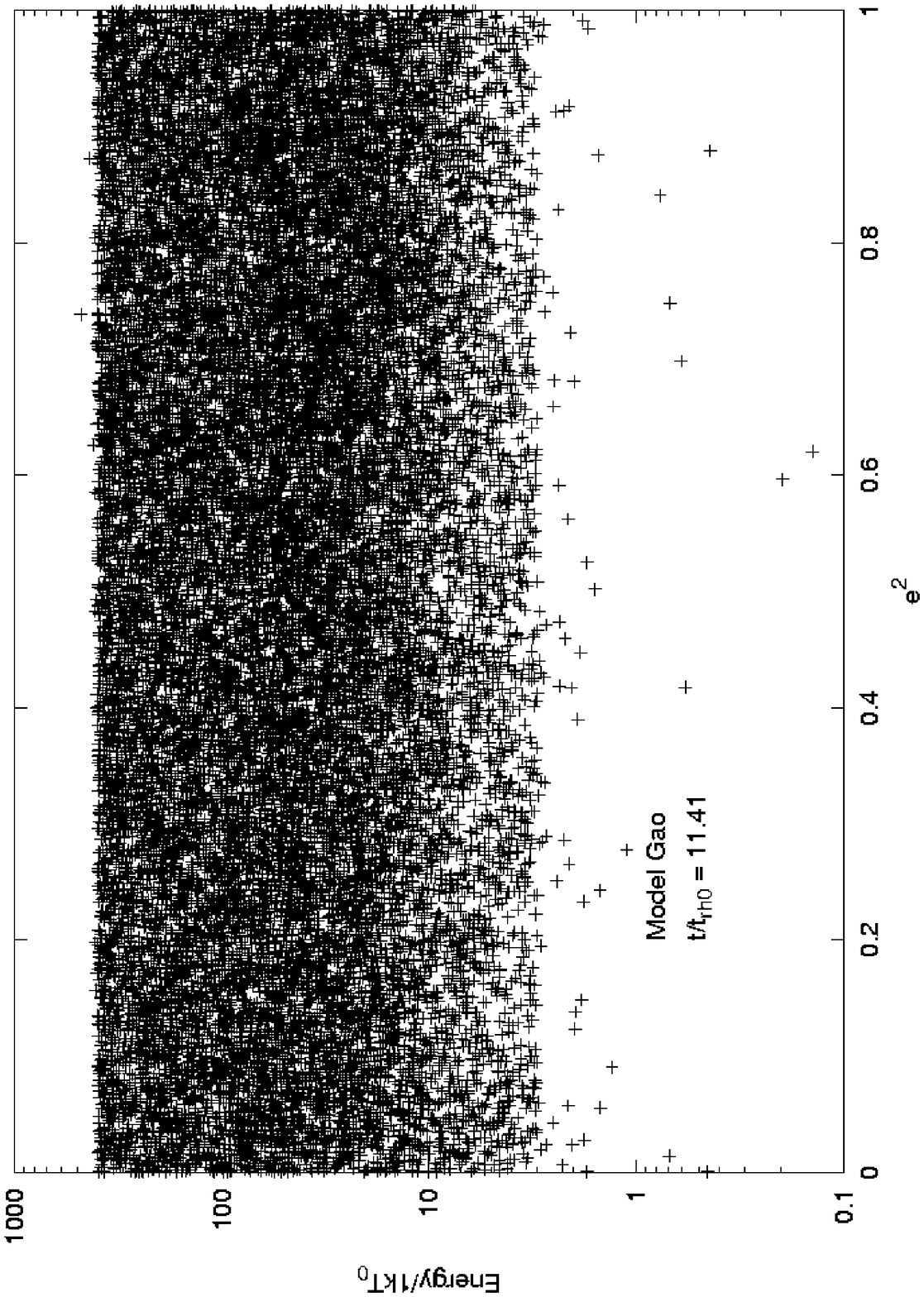}
\includegraphics{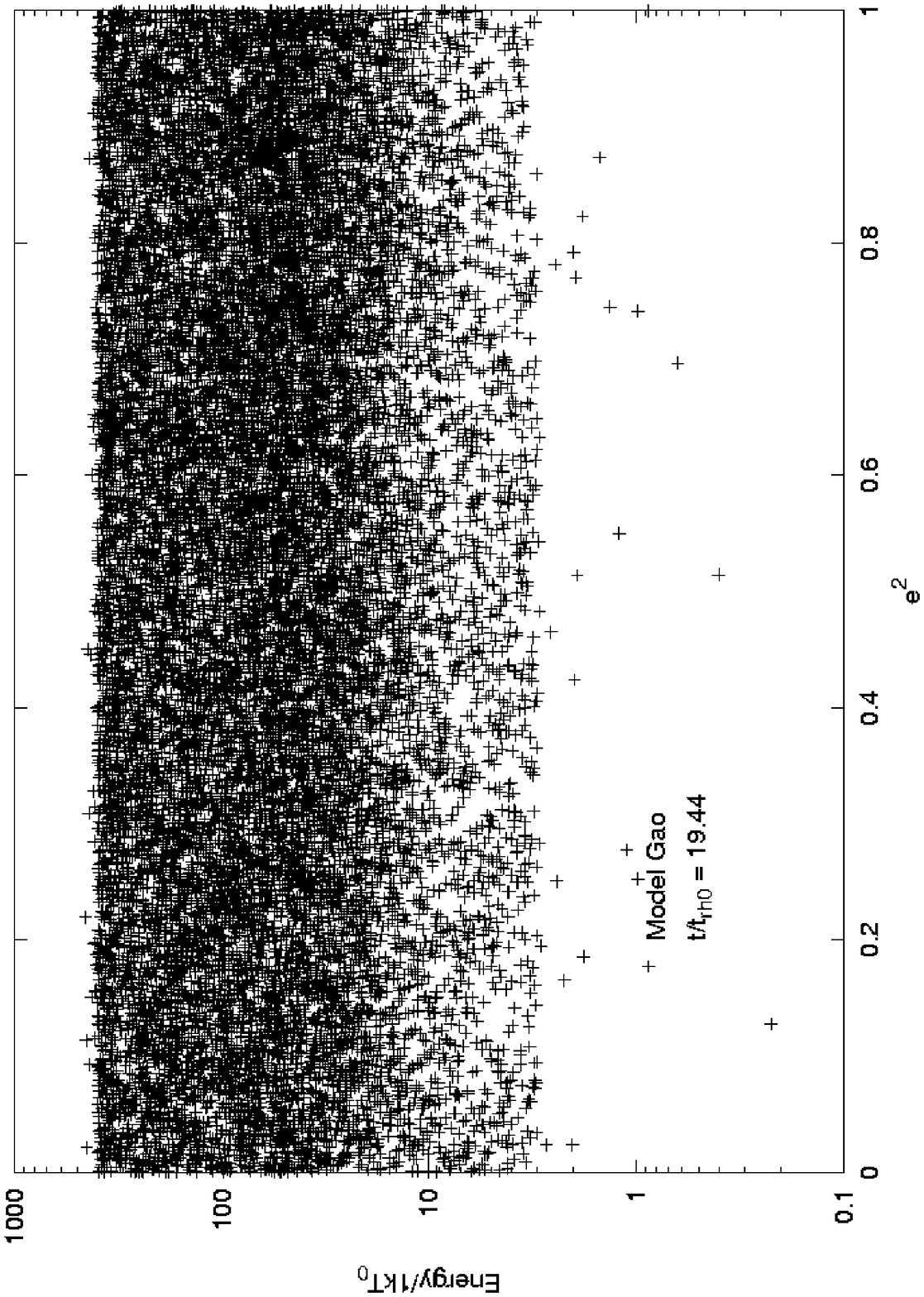}
\includegraphics{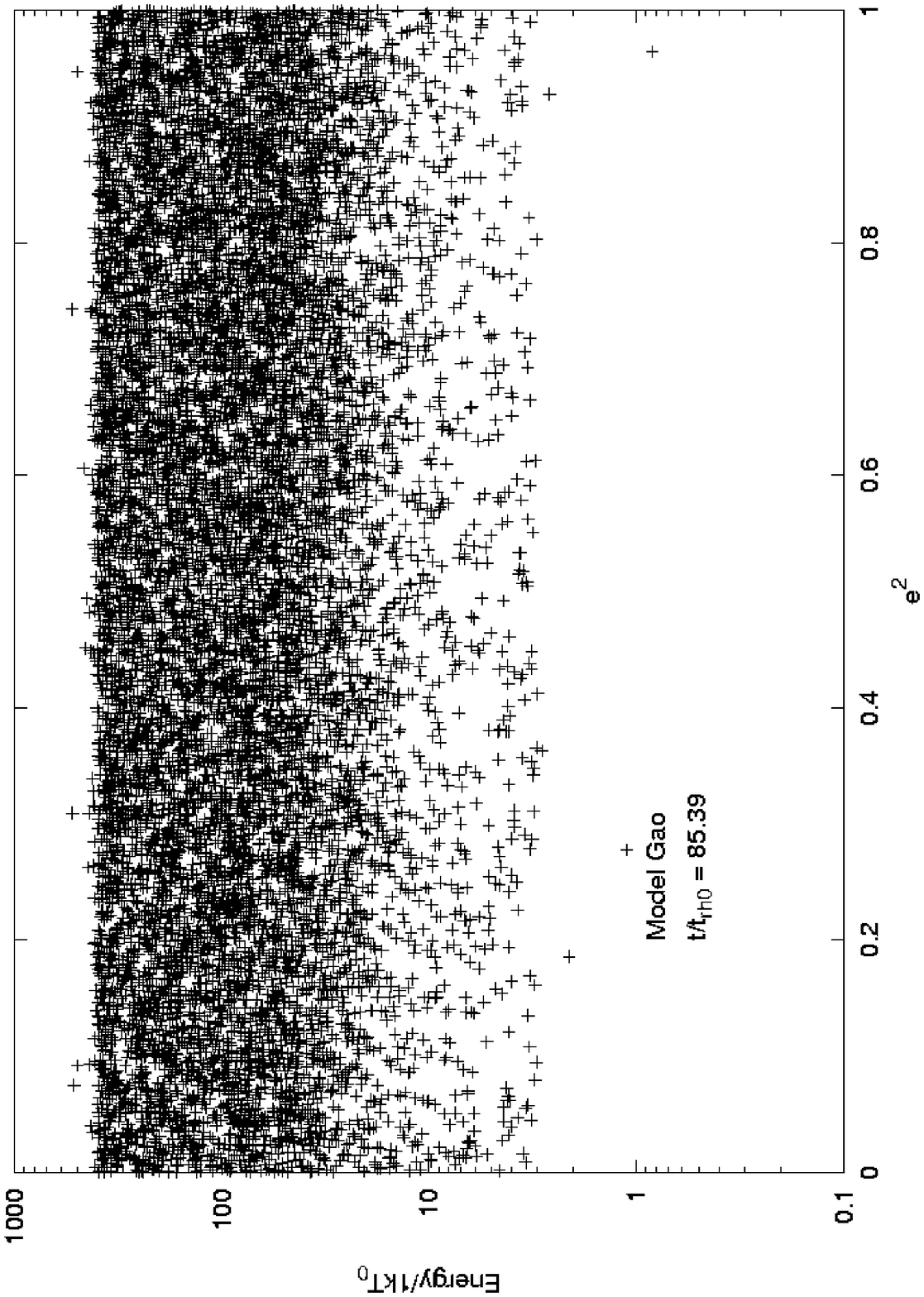}
\includegraphics{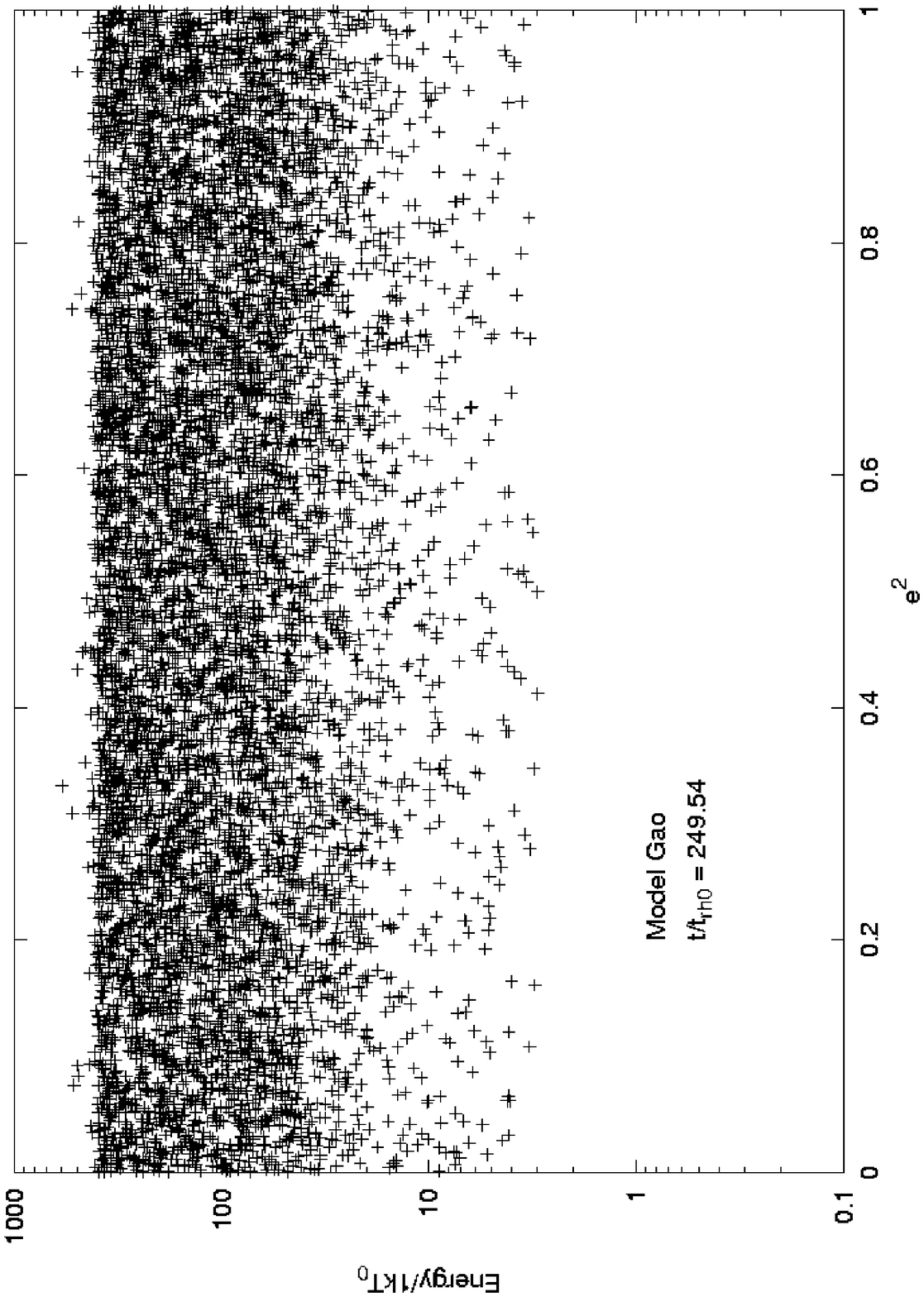}
\includegraphics{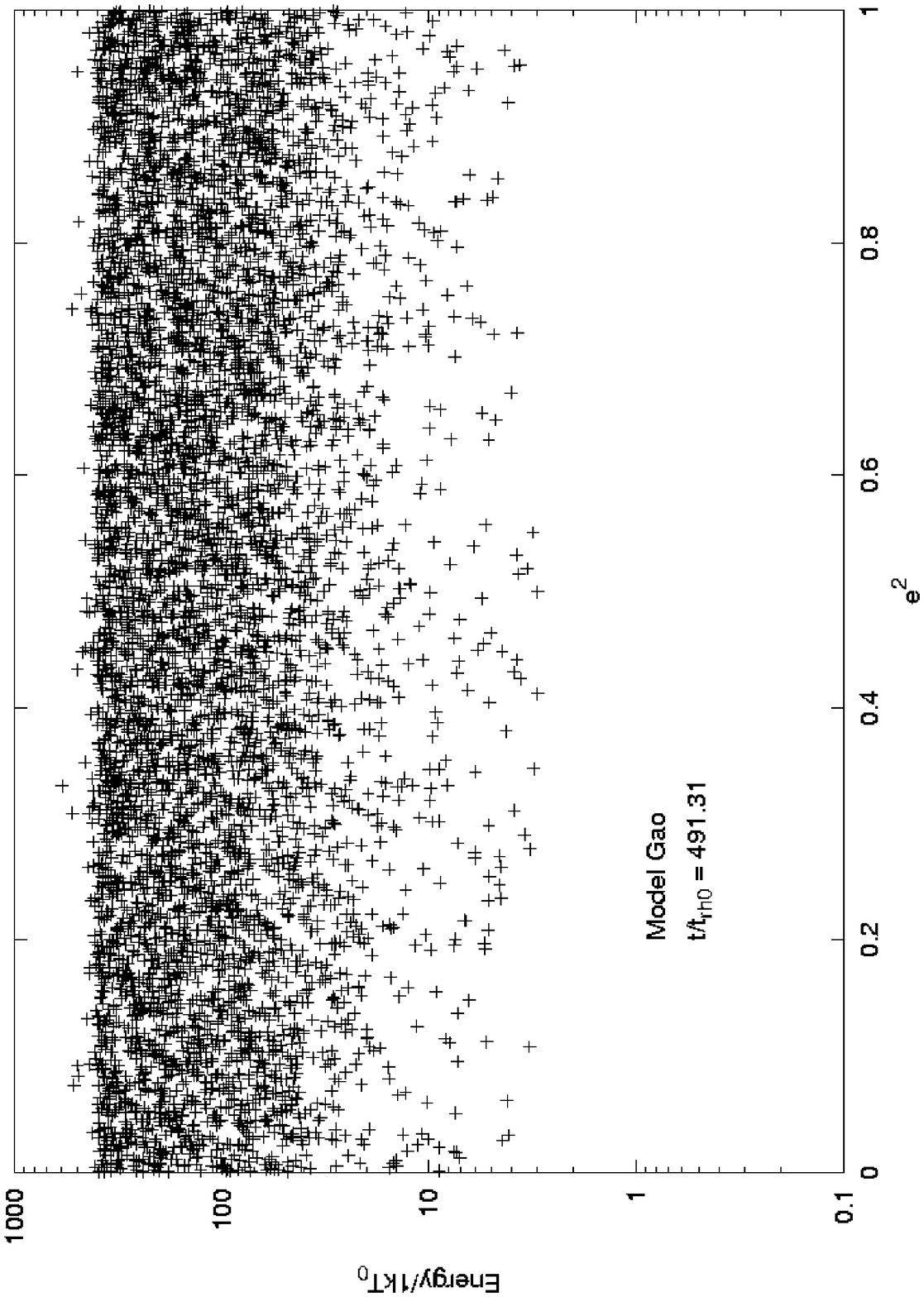}

\caption{Snapshot plots of 2D projection of
 the individual data of each binary (being represented by a cross) onto
 the energy-eccentricity squared plane. Binary binding energy is expressed in
 initial $kT$. Each plot represents different time in units of the initial
 half-mass relaxation time; top left 0, top right 85.39, middle left 11.41,
 middle right 249.54, bottom left 19.44, bottom right 491.31. The selection
 of times is the same as indicated by the arrow in Fig.~\ref{f5}.}
 \label{f6}
\end{figure*}

 \subsection{Gao's Runs}

 Second, we follow our main task, to model a large star cluster with many
 primordial binaries. In order to compare with the previous results of Paper II
 and Gao et al. (1991) we use exactly the same initial conditions, i.e.
 a cluster of 270.000 single stars and 30.000 binaries, both distributed
 in a Plummer's model density distribution with constant density ratio between
 binaries and single stars (for more details see the cited papers).

 All binaries are set up initially
 with a so-called thermal eccentricity distribution,
 where orbital eccentricities are homogeneously distributed as a function
 of $e^2$. Binding energies are distributed logarithmically homogeneous
 between 3 and 400 $kT$, and
 binaries and single star densities are initially obtained from a Plummer model.
 This is analogous to Paper II, except for the eccentricities, which did not
 matter there.

\begin{figure}
 \psfig{figure=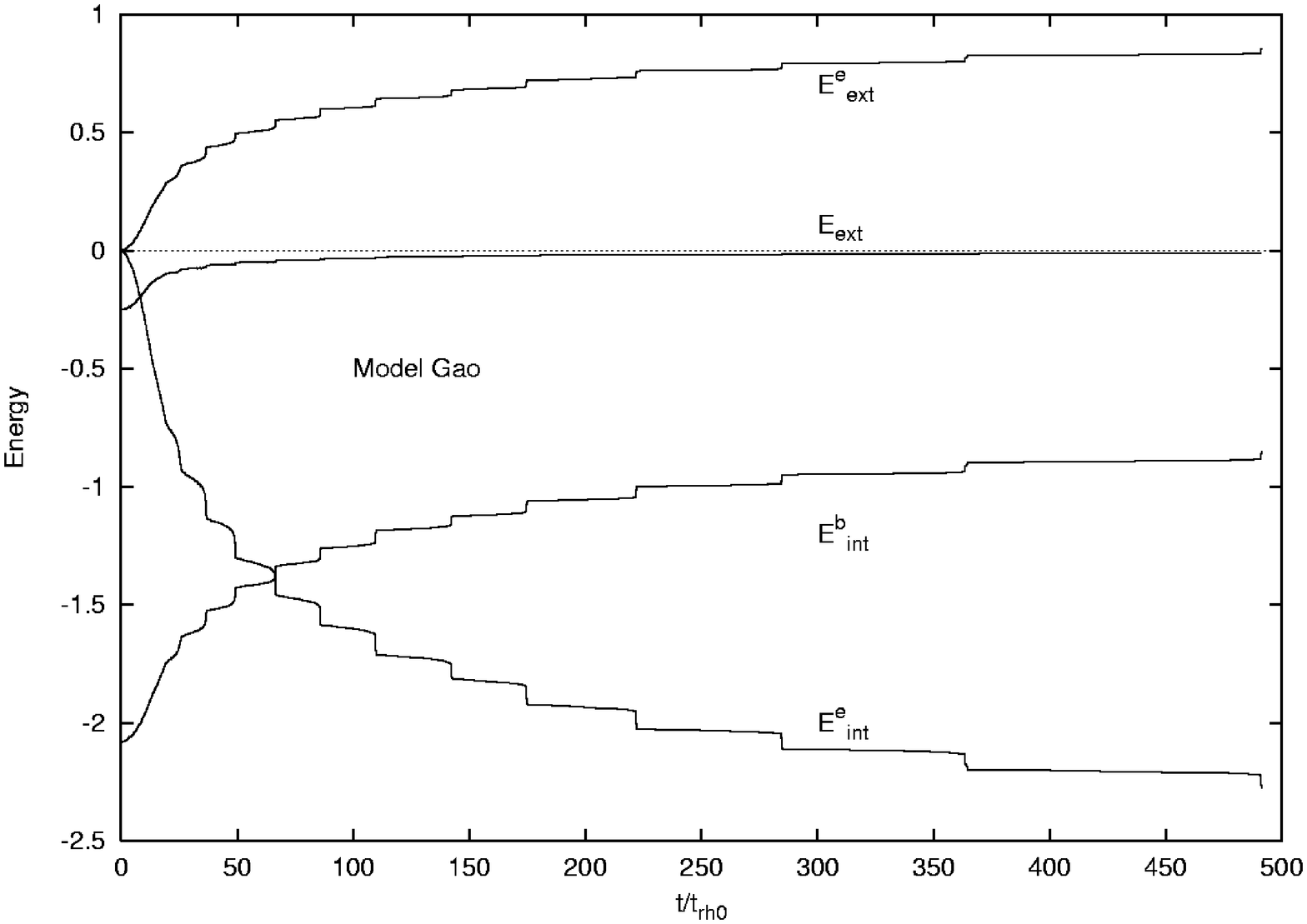,height=6.5cm,width=8.5cm,angle=-90}
 \caption{Energy balance as a function of time. Total internal
 energy of the binaries remaining bound in
 the system ($E^{\rm b}_{\rm int}$), total internal energy of those binaries
 which escaped ($E^{\rm e}_{\rm int}$), total
 external (i.e. potential plus translational) energy of
 all objects (singles and binaries) remaining bound
 in the system ($E^{\rm b}_{\rm ext}$) and escaping
 ($E^{\rm e}_{\rm ext}$).}
 \label{f7}
 \end{figure}

\begin{figure}
 \psfig{figure=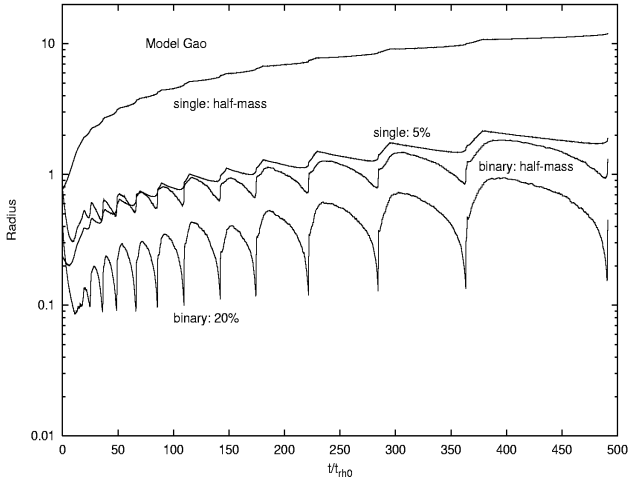,height=6.5cm,width=8.5cm,angle=-90}
 \caption{Evolution of Lagrangian radii containing $50\%$ and $5\%$ of the
 mass of single stars and $50\%$ and $20\%$ of the mass of
 binaries as a function of time (scaled by the
  initial half-mass relaxation time).}
 \label{f8}
 \end{figure}

\begin{figure}
 \psfig{figure=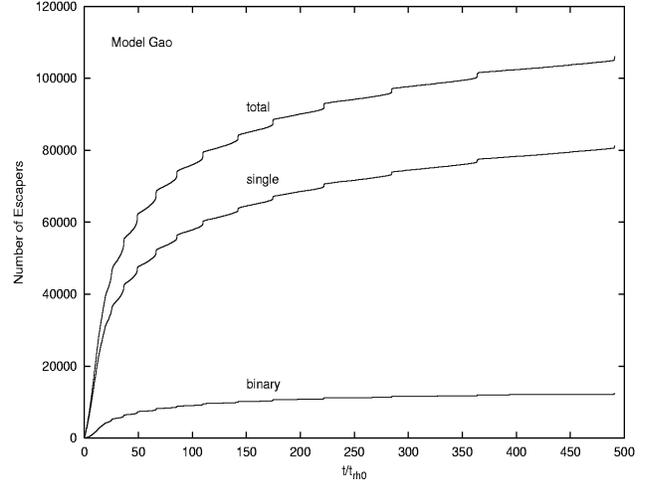,height=6.5cm,width=8.5cm,angle=-90}
 \caption{Number of escaping single stars, binaries and the total number
 of escapers as a function of time (scaled by the
  initial half-mass relaxation time).}
 \label{f9}
 \end{figure}

 One of the main differences from Paper II is that there is no
 quasi-steady binary burning phase before core collapse anymore (compare
 Figs.~\ref{f4} and \ref{f5} as compared with Figs.~19 and 21 of Paper II, respectively). 
 Such a
 quasi-steady state, as it was also seen by Gao et al. (1991), is not
 seen anymore here, because binary-binary interactions have many more
 degrees of freedom here. While in Paper II and in
 Gao et al. (1991) the only assumed outcome of a binary-binary encounter
 is the hardening of the hard binary and destruction of the weaker binary,
 here encounters between moderately hard binaries can yield different and
 non-standard results. Both binaries can survive an encounter if their
 interaction is not too strong, for example. We observe that the total amount of
 internal energy in binaries and the total number of escapers is significantly
 smaller as compared to Paper II (see Figs.~\ref{f7} and ~\ref{f9}), while at
 the same time we have more
 binaries remaining in the system, partly in the outer zones.
 In this work binaries are not as much
 centrally concentrated (see Fig.~\ref{f8}) as compared to Paper II (see there Fig.~26).
 What is not plotted here is that (again compared to Paper II) there is a much
 larger number of binaries surviving in the outermost parts of the cluster.
 We expect that most of these binaries, however, will be lost from the cluster if
 there is a tidal field from the mother galaxy taken into account, which is a
 subject of our ongoing work. In conclusion,
 we think that any difference in the results of Paper II and this work can be ascribed to
 a smaller rate of binary destruction by four-body encounters here.

 Now we turn to the discussion of the detailed eccentricity evolution of our binaries,
 for which there is no counterpart in Paper II.
 Fig.~\ref{f6} shows, that
 during the entire evolution the binary density is constant as a function
 of $e^2$, the squared eccentricity. This is consistent with the older
 predictions by Hills (1975a,b), McMillan et al. (1991) and
 Sigurdsson \& Phinney (1993) that the
 eccentricity distribution in a large number of encounters remains thermal,
 with an average of $\langle e \rangle = 2/3$. However as a note to take care
 in interpreting our Fig.~\ref{f6}, which was obtained from a pure point mass
 model, in a real star
 cluster including stellar and binary evolution, the nice thermal eccentricity
 distribution will not prevail, even on short time scales. Looking for example
 in Kroupa (1995), Fig. 2, top panel, one can see that in pre-main sequence
 evolution high eccentricities of binaries are wiped out by what the author
 calls eigenevolution. For orientation, a circular binary with semi-major axis of
 1 astronomical unit ($\log P \approx 2.56$ in Kroupa's figure) would correspond
 roughly to a 10 $kT$ binary in our models. In other words, in a cluster involving
 stellar evolution binaries with 10 $kT$ and eccentricity $e>0.6$ would be
 subject to strong eigenevolution and may thus disturb the thermal eccentricity
 distribution. It is worth to note that binaries with binding energy equal to about
 400 $kT$ have semi-major axis roughly equal to the solar radius (in units
 used in our simulations), so they will be
 subject to merge before interacting with other stars.

 Figs.~\ref{f6} show in six panels the evolution of our initially 30.000 binaries
 in binding energy -- eccentricity space. They all start equally distributed in
 a logarithmic scale of $E_b$ and $e^2$. While as expected more and more binaries
 are disrupted an interesting result is that there is no tendency to accumulate
 anywhere in eccentricity--energy space. We always have binaries of any eccentricity,
 with no obvious eccentricity binding energy correlation, even in the late
 phases of the evolution. This is very different from the radius energy relationship,
 which was plotted in Paper II in a similar way, where there was a correlation
 in the sense that the core does not have weak binaries, because they are destroyed.
 The stationary thermal eccentricity distribution of binaries is consistent with
 the previous scattering studies of Hills (1975a,b), McMillan et al. (1991)
 and Sigurdsson \& Phinney (1993),
 but here for the first time confirmed in combination with an evolving binary
 population in a strongly evolving cluster.

\begin{figure}
 \psfig{figure=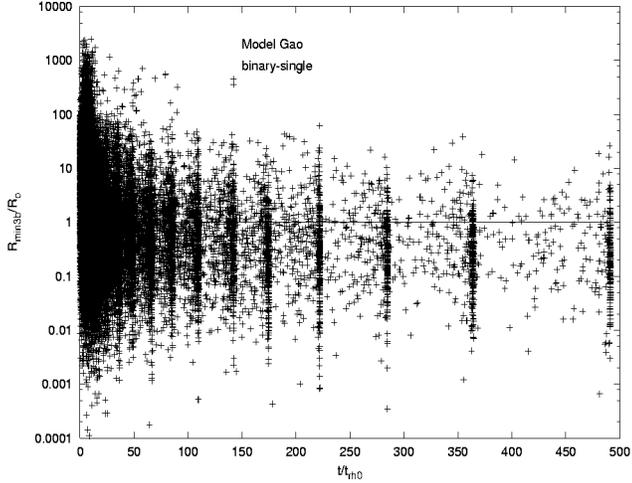,height=6.5cm,width=8.5cm,angle=-90}
 \caption {Minimum distance during binary-single interaction (in solar unit)
  as a function of time (scaled by the
   initial half-mass relaxation time).}
 \label{f10}
 \end{figure}

\begin{figure}
 \psfig{figure=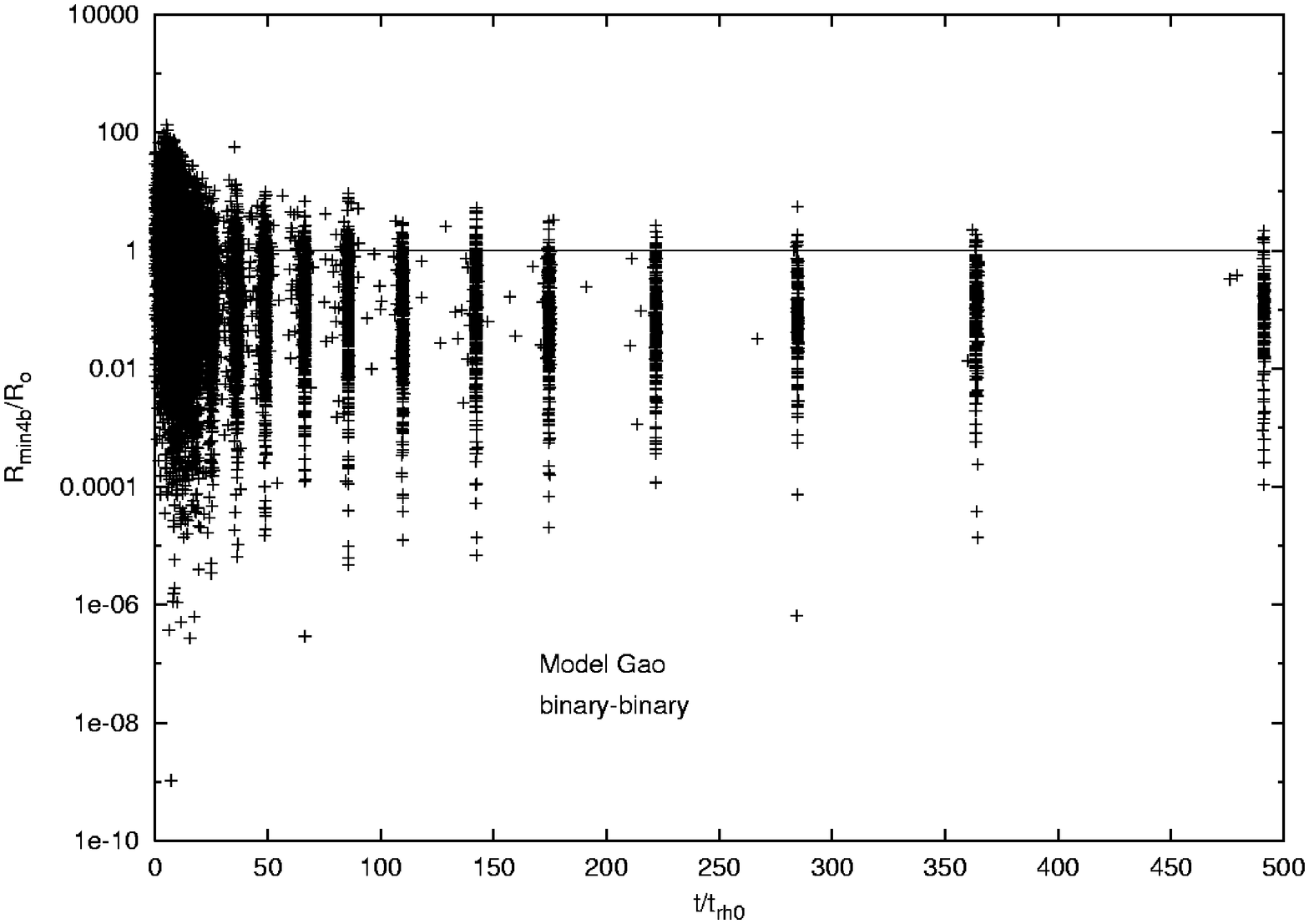,height=6.5cm,width=8.5cm,angle=-90}
 \caption{Minimum distance during binary-binary interaction (in solar unit)
  as a function of time (scaled by the
   initial half-mass relaxation time).}
 \label{f11}
 \end{figure}

 Following every encounter in detail using our numerical integration scheme
 gives us the possibility to check for closest encounter distances and
 possible mergers. Different from any other result discussed in this paper
 we have to fix the physical scales if we want to give meaningful
 results for minimum encounter distances in units of stellar or solar radii.
 For the scaling here we have assumed that the scaling radius of our
 initial Plummer model is 1.0 pc and that fixes everything in combination
 with the assumption that one particle (one star) has one solar mass.
 With our total mass of 330.000 solar masses this results in rather
 large central densities for our initial model, comparable only to the most concentrated
 globular or massive clusters near the galactic centre.
 Therefore the discussion of ``merging'' in any of
 the following figures should be seen as exemplary, since the rate of
 merging depends on the assumed scaling in the physical units. Any other quantities though,
 like the relative distribution of minimum encounter distances in
 scaled  $N$-body units or scaling behaviour of cross sections is
 independent of the selected physical units. One should also bear in
 mind, that we do not merge any objects in our dynamical model,
 so it is possible that a very tight binary causes a ``merger'' event
 in our simulations and it is counted as a merger several times (see Fig.
 ~\ref{f10} and \ref{f11}). In a subsequent stage of this work merging will be included.

 In Figs.~\ref{f10} and \ref{f11} we show for each individual three- or
 four-body interaction, whether the minimum distance occurring during the
 interaction is larger than one solar radius (greater than unity) or
 vice versa. In the late stages the signature of density maxima in
 gravothermal oscillations can be seen, and in the beginning there is
 a lot of events of all kinds due to the large number of binaries present.
 We prefer to show this rather primitive plots nevertheless, because even
 on that level one can easily by eye estimate that one half of all
 three-body encounters, and some three quarters of all four-body encounters
 would probably lead to a merger if the stars would have solar mass and radius (but
 keep in mind discussion in the previous paragraphs about scaling simulations
 units to the physical units).
 In the late evolutionary phase with gravothermal oscillations nearly
 all four-body encounters lead to a merger of solar-type stars.

 \begin{figure}
 \psfig{figure=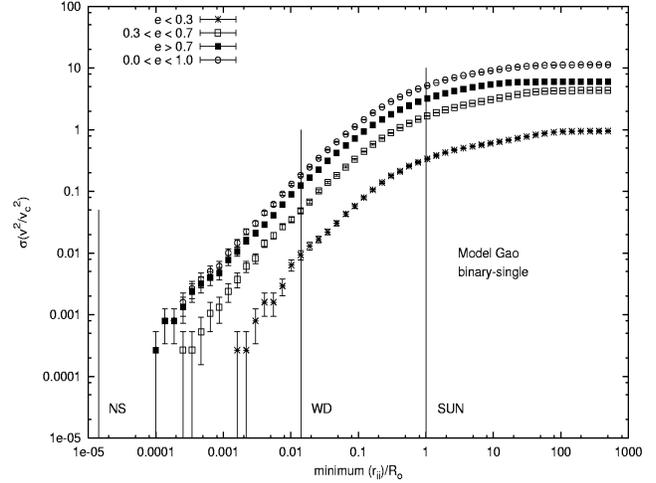,height=6.5cm,width=8.5cm,angle=-90}
 \caption{The cumulative cross section for pairwise close approach during
 binary-single interactions. The separation is scaled to the solar radius.
 The cross section is normalized to the geometrical cross section of the
 binary and corrected for gravitational focussing. The normalization factor
 is properly averaged over all interactions (see for details in the text.}
 \label{f12}
 \end{figure}

\begin{figure}
 \psfig{figure=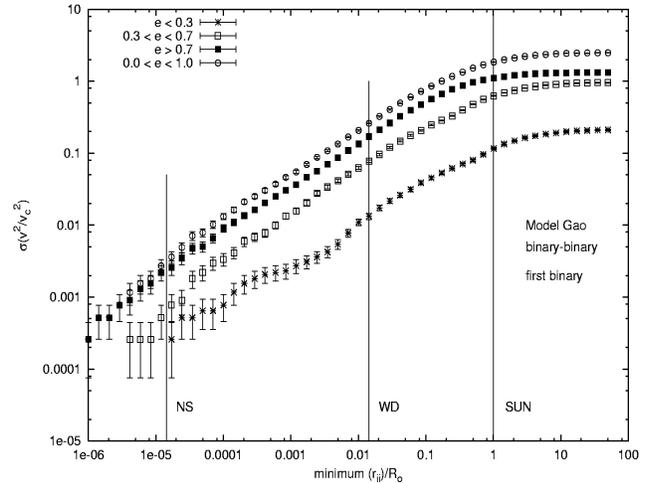,height=6.5cm,width=8.5cm,angle=-90}
 \caption{The cumulative cross section for pairwise close approach during
 binary-binary interactions. The separation is scaled to the solar radius.
 The cross section is normalized to the geometrical cross section of the
 binaries and corrected for gravitational focussing. The normalization factor
 is properly averaged over all interactions (see for details in the text.)}
 \label{f13}
 \end{figure}

 In Figs.~\ref{f12} and \ref{f13} we show how the initial eccentricities of
 the binary (the harder binary in case of four-body encounters) influence
 the probability of a close encounter distance, again using the minimum
 distance scaled to solar values. We show four curves; the one
 labelled $0.0 < e < 1.0 $ gives the cumulative cross section for all
 encounters. The three other curves
 show the same cross section for initially more circular orbits ($e<0.3$),
 highly eccentric ($e> 0.7$) and moderately eccentric ($0.3 \le e \le 0.7 $)
 orbits. The cross section for merging is nearly one order of magnitude
 larger for encounters which involve highly eccentric binaries as compared to
 those where there are only less eccentric or circular binaries. While this
 is relevant for stars with solar radii, it can be seen that if we look for
 much smaller minimum encounter distances, as they would correspond e.g. to
 white dwarf or neutron star mergers, that for those the merger cross
 sections are totally dominated by mergers obtained from highly eccentric
 binaries. In other words any merger of white dwarfs, neutron stars or
 other compact objects is highly likely to have originated from a very eccentric
 binary (not that surprising indeed).

 One can also see in Figs.~\ref{f12} and \ref{f13} that the cumulative
 cross section $\sigma$ scales approximately with $d$ for three-body and
 with $\sqrt{d}$ for four-body encounters, where $d$ is the scaled
 minimum encounter distance as in the figures. Hut \& Inagaki (1985)
 discuss that a linear scaling is obtained for very small $d$ by pure
 geometrical approach plus gravitational focusing, while the flattening
 of $\sigma$ for larger $d$ occurs due to the importance of resonant
 encounters, and they derive an empirical law of $\sigma\propto d^{0.4}$,
 which is in fair agreement with our findings. Our results tell us
 in connection with the discussion in Hut \& Inagaki (1985) that most
 of our four-body encounters are indeed resonant, while the three-body
 encounters include many non-resonant encounters. This is a consequence
 that we limit $p_{\rm max}$ to $10 a_1$ in the case of four-body
 encounters as discussed above.

 \begin{figure}
 \psfig{figure=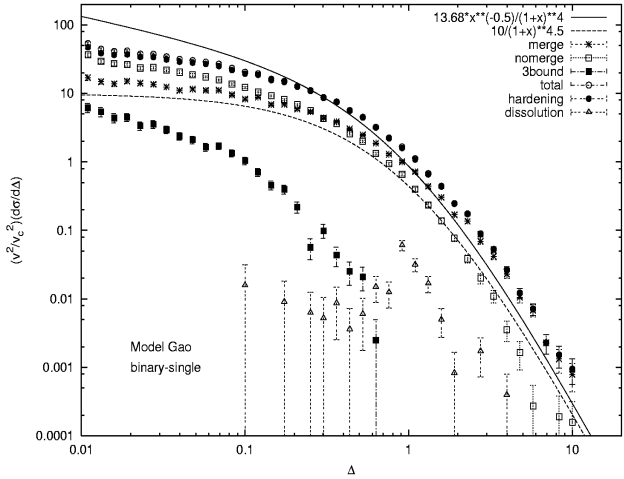,height=6.5cm,width=8.5cm,angle=-90}
 \caption{Differential cross sections for relative binding energy changes
 ($\Delta$)  as a function of absolute value of $\Delta$, for binary single
 star interactions. The
 differential cross section is normalized to  the geometrical cross section
 of the binary and corrected for gravitational focussing. The normalization
 factor is properly averaged over all interactions (see text for details).
 The continuous line indicates the Spitzer's analytical estimate, the dashed
 line is for Heggie's estimate. The following symbols are for: mergers - star,
 no mergers - open square, 3-body bound - filled square, total - open circle,
 hardening - filled circle and dissolution - open triangle.}
 \label{f14}
 \end{figure}

\begin{figure}
 \psfig{figure=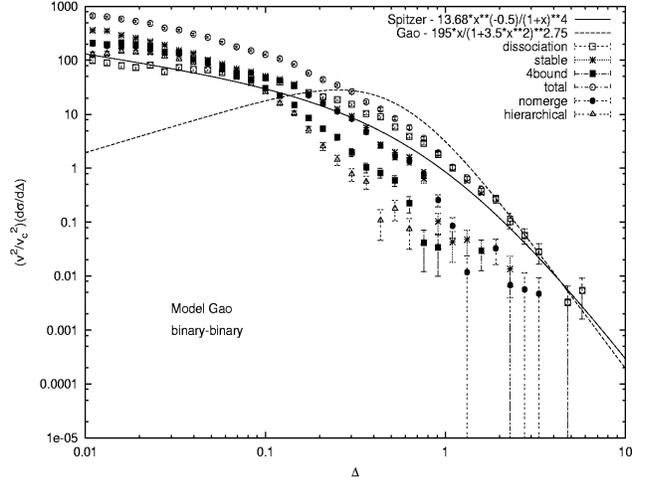,height=6.5cm,width=8.5cm,angle=-90}
 \caption{Differential cross sections for relative binding  energy changes
 ($\Delta$) as a function of absolute value of $\Delta$, for binary binary
 interactions. The differential
 cross section is normalized to the geometrical cross section of the binary
 and corrected for gravitational focussing. The normalization factor is properly
 averaged over all interactions (see text for details).
 The continuous line indicates the Spitzer's analytical estimate, the dashed
 line is for Gao's estimate. The following symbols are for: stable - star,
 dissolution - open square, 4-body bound - filled square, total - open circle,
 no merge - filled circle and hierarchical - open triangle.}
 \label{f15}
 \end{figure}

 In Figs.~\ref{f14} and \ref{f15} we show, in the standard way, empirical
 normalized differential cross sections as a function of $\Delta=(E_b^\prime - E_b)/E_b$, the
 normalized energy change of the harder binary ($E_b^\prime$, $E_b$ binding energies
 after and before the encounter of the binary; in case of dissolution or fully bound
 configurations we use $E_b^\prime = 0$). The normalization is done here
 over $\pi a^2$, where $a$ is the average semi-major axis of all binaries involved
 in the encounters. For three-body encounters
 (Fig.~\ref{f14}) we find a fairly good agreement of the nearly entire differential cross
 section with Spitzer's expression

 \begin{equation}
 {v^2 \over v_c^2} {d\sigma\over d\Delta} = { 13.86 \over \Delta^{1/2}(1+\Delta)^4}
 \end{equation}
 which is plotted into the figure for comparison. For hard encounters there is
 also little difference to Heggie's cross section
 $\propto (1+\Delta)^{-4.5}$. Notably, however, for
 very small energy changes, our empirical cross sections are different
 from Spitzer's expectation. It is not surprising that such differences occur here,
 because in our real cluster model there is only a limited coverage of phase space
 for all encounters with small energy changes (large impact parameters), different
 from artificial series of three-body experiments.
 In our real cluster time and $\pmax$ are limited, so we have a smaller number of
 events here then in a theoretical simulation experiment of three- and four-body
 encounters. While this could be seen as an error from a standpoint of strict theoretical
 determination of cross sections we stress that it is our interest to derive statistical
 expectations to what happens to binaries in real clusters, which is truly reflected in
 our models. In order to avoid too many plots we have depicted in the plots multiple
 representations of the same data. The entire cross section includes all events
 (note that we have used ``total'' in the plot as a label for that, which is still
 a differential cross section, but including all possible reaction outcomes).
 Our cross sections are then subdivided in two different ways: first differentiating
 between
 merging and non-merging events (here for one solar radius as minimum distance),
 second differentiating between physical end states such as three bound, hardening
 and dissolution (in case of four-body encounters: stable hierarchical triple).

 First discussing three-body cross sections in Fig.~\ref{f14} we see that
 three bound and dissolution have a very small probabilities, the latter occurs preferentially
 in strong encounters (high energy change of binary), while the
 second has only a chance for small (negative) energy changes. Only for small
 $\Delta$ there is some gap between the entire and hardening cross sections, at
 larger $\Delta$ the difference between the two is statistically insignificant.
 Hence, most strong three-body encounters result in a hardening, except at small
 energy changes, a well-known result which demonstrates the reliability of
 our methods. Since dissolution
 plays no significant role statistically (by a margin of more than two orders
 of magnitude in the measured cross section at small $\Delta$), most non-hardening
 events are softening (not
 plotted separately). The case three bound can only occur at negative $\Delta$,
 so it cannot compete with the three other cases at the same $\Delta$. We have plotted
 it nevertheless in the same figure (using the absolute value of $\Delta$).

 Merging occurs at high $\Delta$ (strong hardening events) with higher probability
 than at low $\Delta$, but for all $\Delta$ the difference in cross section is less
 than an order of magnitude. This already demonstrates that there must be another
 factor determining close approach, which is the initial eccentricity of binaries
 in encounters, as was seen above for the total cross sections as a function of
 minimum distance, for different initial eccentricities (Figs.~\ref{f12} and \ref{f13}).

 In the case of differential cross sections for four-body encounters there is
 little analytical work to compare with. Our three- and four-body cross sections
 can be compared with the pioneering work of McMillan et al. (1991) for small
 $N$body simulations (see Figs. 13-16 there).
 We also use Gao's analytical cross section (Gao et al. 1991)
 in Fig.~\ref{f15} for comparison. It assumes that each four-body encounter
 results in a dissolution of the weaker binary, and a hardening of the hard
 binary by some fraction of the sum of the initial binding energies. For strong
 encounters, where dissociation dominates the cross section, as seen in Fig.~\ref{f15},
 Gao's cross section is not bad. For $\Delta < 1$ we have a competition
 between dissociation and stable end configurations (resulting in two surviving
 binaries). At small energy changes formation of bound quadruples and stable
 hierarchical triples is the most probable reaction channel. Since we do
 not yet follow the evolution of the bound four-body systems we cannot
 say anything definite about their future. But many of them would possibly
 just be stripped off their fourth star and result in triples, where again
 there would be a considerable number of stable triples.

 \begin{figure}
 \psfig{figure=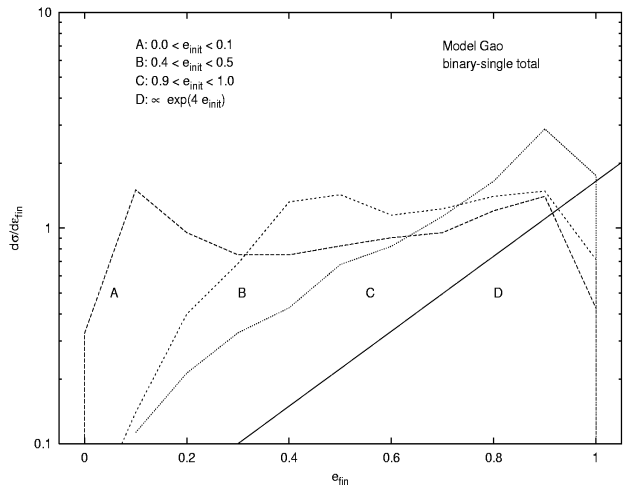,height=6.5cm,width=8.5cm,angle=-90}
 \caption{Differential cross sections for eccentricity changes as a function
 of final eccentricity, for all binary-single interactions. Curves
 labeled by A, B, C are for small initial eccentricities, medium initial
 eccentricities and large initial  eccentricities, respectively. Line
 labeled by D is the exponential fitting done by eye.}
 \label{f16}
 \end{figure}

 \begin{figure}
 \psfig{figure=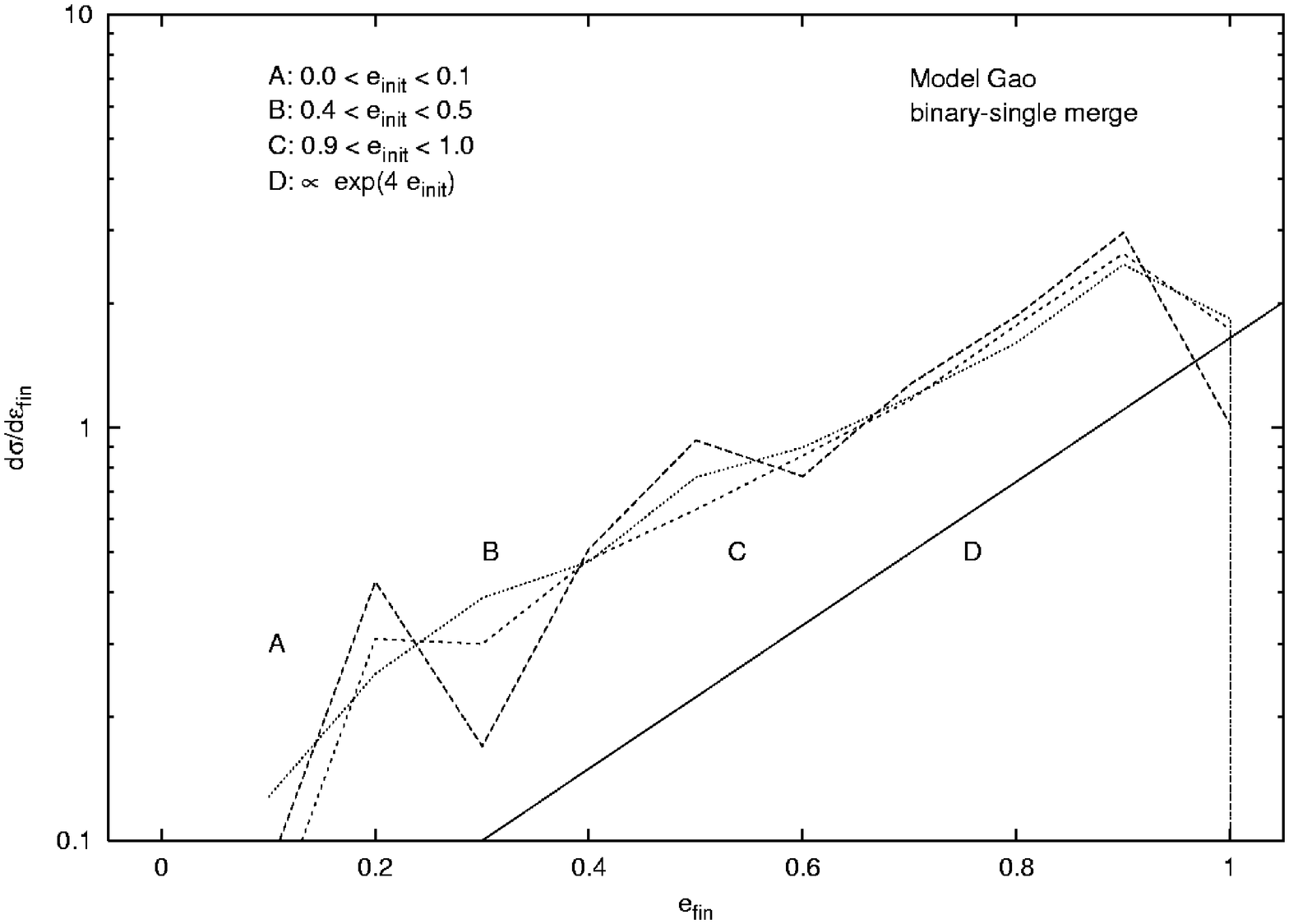,height=6.5cm,width=8.5cm,angle=-90}
 \caption{Differential cross sections for eccentricity changes as a function
 of final eccentricity, for binary-single interactions with mergers. Curves
 labeled by A, B, C are for small initial eccentricities, medium initial
 eccentricities and large initial  eccentricities, respectively. Line
 labeled by D is the exponential fitting done by eye.}
 \label{f17}
 \end{figure}

 \begin{figure}
 \psfig{figure=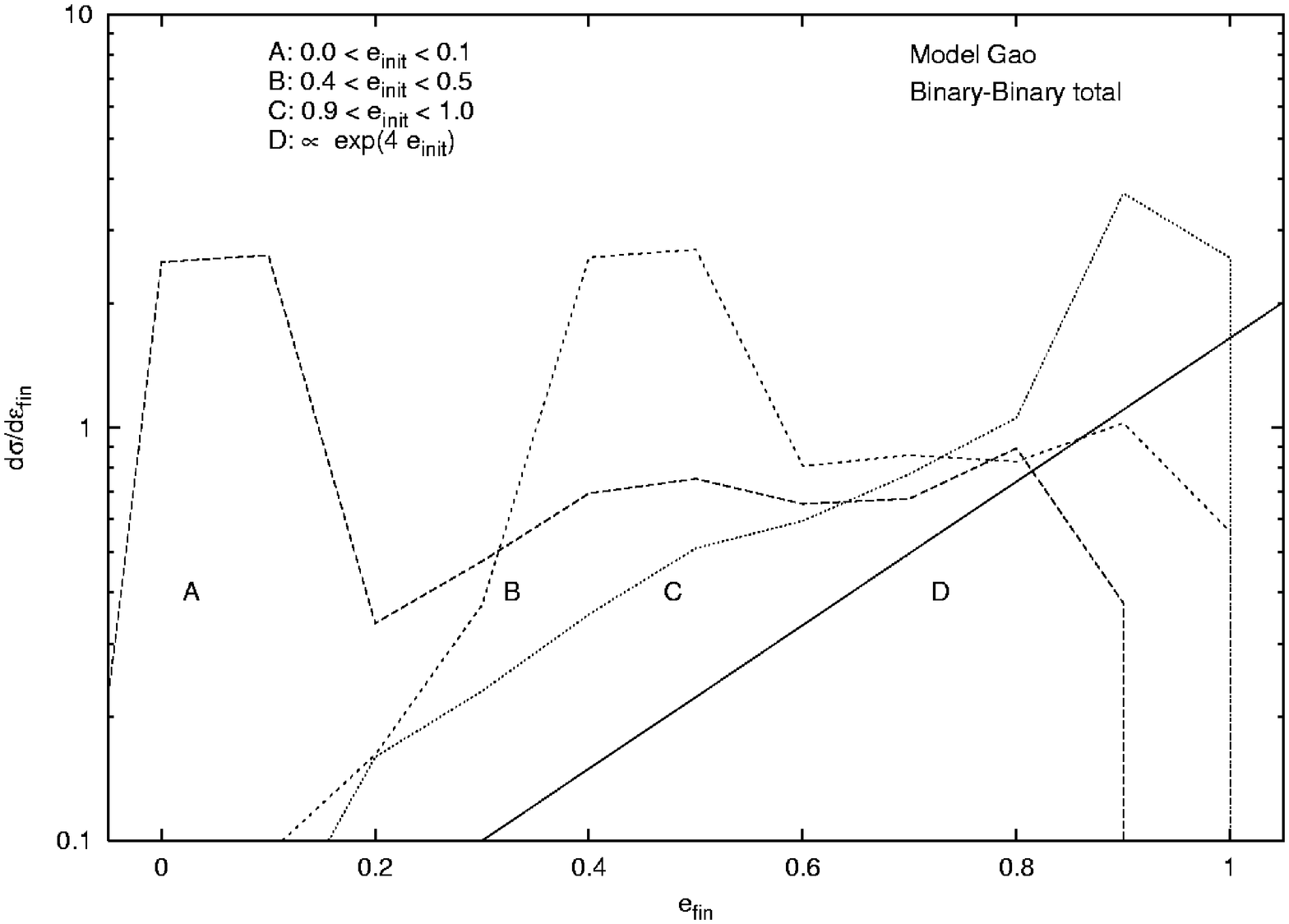,height=6.5cm,width=8.5cm,angle=-90}
 \caption{Differential cross sections for eccentricity changes as a function
 of final eccentricity, for all binary-binary interactions. Curves
 labeled by A, B, C are for small initial eccentricities, medium initial
 eccentricities and large initial  eccentricities, respectively. Line
 labeled by D is the exponential fitting done by eye.}
 \label{f18}
 \end{figure}

 \begin{figure}
 \psfig{figure=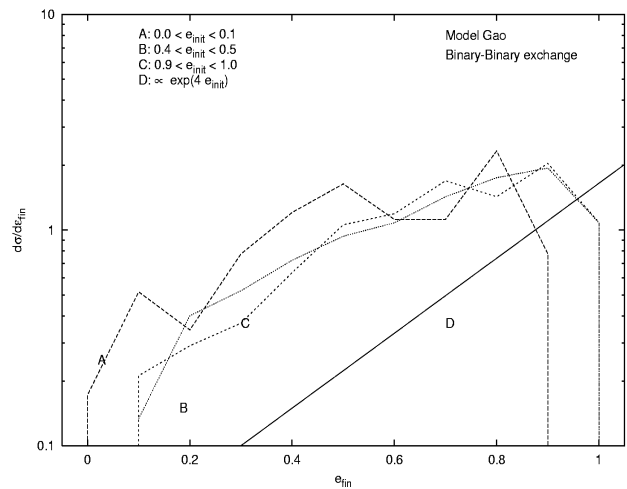,height=6.5cm,width=8.5cm,angle=-90}
 \caption{Differential cross sections for eccentricity changes as a function
 of final eccentricity, for binary-binary interactions with exchange of stars.
 Curves  labeled by A, B, C are for small initial eccentricities, medium initial
 eccentricities and large initial  eccentricities, respectively. Line
 labeled by D is the exponential fitting done by eye.}
 \label{f19}
 \end{figure}

  Finally we discuss a new kind of differential cross sections which could be obtained
 from our data as far as we know for the first time. These are differential cross
 sections for eccentricity changes as a function of the final eccentricity. These results,
 shown in Fig.~\ref{f16} for three-body encounters, give the differential cross section for a
 certain final eccentricity to occur. For initially nearly circular orbits (curves labelled A,
 initial eccentricity $e_{\rm init}< 0.1$) we find that all final eccentricities after
 a three-body encounter occur with nearly equal probability. If there is already some initial
 eccentricity (see curves labelled B, for $0.1 < e_{\rm init} < 0.5 $ the probability
 to reach any higher eccentricity is approximately constant, while the chance to go
 back to a less eccentric orbit decays exponentially. This is even more pronounced
 for initially highly eccentric binaries ($e_{\rm init} > 0.9$): here eccentricity
 remains high, and there is an exponential law $\propto \exp(4 e_{\rm fin})$ which could
 be fitted to describe the decreasing differential cross section to reach smaller
 eccentricities.

 Turning to four-body encounters the entire cross section (labelled ``total'' in the
 figure) depicted in Fig.~\ref{f18} shows at first glance a different behaviour.
 It seems there is a tendency for most binaries to keep their initial eccentricity -
 the differential cross section has a maximum at $e_{\rm fin} \approx e_{\rm init}$!
 However, this is not really the case, rather we have a bimodal distribution depending
 on whether we look at strong encounters or weak encounters. If we discriminate
 strong encounters by our merger or exchange criterions (minimum distance smaller than 
 solar radius or one of the binary component is exchanged during the interaction),
 we see in Figs.~\ref{f17} and \ref{f19} that for a strong encounter
 the initial eccentricity is ``forgotten'' in the sense
 that all differential cross sections have a maximum at high final eccentricities and
 decay again with the characteristic $\propto \exp(4 e_{\rm fin})$ law seen already
 in three-body encounters. The maxima at $e_{\rm fin} \approx e_{\rm init}$ seen in
 Figs.~\ref{f16} and \ref{f18} could be ascribed totally to weak encounters (fly-bys) in which
 there is no strong interaction and hence no strong eccentricity change
 (compare Heggie \& Rasio 1996). Such fly-by feature in the eccentricity cross section
 is less pronounced in the case of three-body encounters because we do not extend
 $\pmax$ to large enough values (see discussion above).
 For four-body encounters $\pmax$ depends on the semi-major axis of the second
 binary which is usually much larger than for the first one. So we can expect much 
 more fly-by interactions than in the three-body encounters.

 \section{Conclusions and discussion}

 We have performed a fully self-consistent numerical simulation of a point-mass star
 cluster consisting of 270.000 single stars and 30.000 hard binaries initially.
 All binaries have thermal eccentricity distribution and a logarithmically constant
 binding energy distribution at the beginning. All stars, being member of a binary
 or not, have the same mass. The cluster evolution of single stars and binaries is
 followed using the stochastic Monte Carlo model described in Spurzem \& Giersz (1996, Paper I)
 and Giersz \& Spurzem (2000, Paper II). Instead of using statistical cross sections
 to determine the outcome of close three- and four-body encounters (binary-single star and
 binary-binary scatterings) we use here a direct regularized three- and four-body integration,
 using the methods of Aarseth \& Zare (1974) for three and Mikkola (1983) for four bodies
 (unperturbed motion). The method is tested in comparison with $N$-body modelling of
 star clusters with primordial binaries (Heggie \& Aarseth 1992) and in comparison with
 Paper II. With the new detailed integration of binary interactions we can follow many
 different channels of these encounters, such as dissociation of binaries, weakening
 (instead of only hardening or increase of binding energy), formation of hierarchical
 triples and bound three- and four-body systems. Since we know the orbital elements of
 all binaries we can discuss differential cross sections not only regarding binary
 binding energy changes but also eccentricity changes. Furthermore we are able to monitor
 during encounters distances of closest approach and discuss total cross sections for mergers
 to occur as a function of minimum approach distances. No mergers are considered in our
 models, we just continue point-mass integration. Thus, our data only allows to
 predict some fraction of encounters which would lead to mergers given a critical
 distance of closest approach. Subsequent evolution of the merger remnants as well as
 of the bound triple and quadruple configurations is subject of future work. For the
 case of stable hierarchical triples according to the stability criterion of Mardling \& Aarseth
 (2001) we again discuss the differential cross sections leading to such configurations,
 but do not yet follow the further evolution of triples in our model.

 The main new feature of our improved model as compared to Paper II which only
 used statistical cross sections is that binaries dissolve much quicker in a
 sequence of encounters increasing their eccentricity and subsequent dissolution
 by four-body encounters. The
 time of quasi-stationary binary burning is now very much shorter than before (about 50 $t_{rh}$)
 and gravothermal oscillations follow (Bettwieser \& Sugimoto 1984). We perform a detailed
 statistical analysis of all encounters between binaries and binaries and single stars,
 to determine total and differential cross sections. While this is technically done in
 the standard way, see Hut \& Bahcall (1983), Mikkola (1983), Bacon, Sigurdsson \& Davies (1996)
 an important physical difference is that we model such encounters in the environment
 of a real star cluster evolution with its limitations of phase space available for
 initial conditions of encounters (limited time, limited maximum impact parameter).
 As a consequence we find that most of our empirical cross sections agree very well
 with the expected analytical ones of Spitzer (1987) for three-body and Gao et al. (1991) for
 four-body encounters for strong encounters (large binding energy changes). For weaker
 interactions we find a much more differentiated result, which is important for
 the overall dynamics (energetics) of our cluster.

 Finally, we monitor the eccentricity of all binaries. In all evolutionary stages a so-called
 thermal eccentricity distribution is maintained at all binary binding energies. Nevertheless
 we find that strong encounters (e.g. those leading to merging and dissolution of binaries)
 are predominantly occurring for initially very eccentric binaries. This seems to be a
 contradiction (why are high eccentricities not depleted?), but is explained by our newly
 defined and measured cross sections for eccentricity changes. Close three- and four-body
 encounters all have a high probability (maximum of differential cross section) to yield
 high eccentricities. Having reached high eccentricity such binaries are particularly vulnerable
 to strong encounters which lead to very small minimum distances, large energy changes, and
 consequently also lead to disruptions. We find a new scaling law for the differential
 cross section of eccentricity changes as a function of final eccentricity which scales
 with an exponential $\propto\exp(4 e_{\rm fin})$. This has no theoretical background yet and
 should be considered as purely empirical for the time being.

 Having achieved one more significant step toward a full, self-consistent dynamical modelling
 of rich star clusters with all their binaries (and their internal degrees of freedom) we are
 now planning the next step. By use of our detailed $N$-body integration for close encounters we
 will be now able to vary the masses of the stars (since the unknown cross sections are not
 needed to determine the results), including a mass spectrum. Furthermore, we can then apply
 some recipes to merge stars if they overlap and increase their masses correspondingly. Also
 hierarchical triples should be followed in the future for their further internal evolution
 and perturbations by subsequent encounters with stars and binaries. This will open up paths
 to predict certain types of merger remnants in star clusters, their distribution and frequency
 in time and distance from the cluster centre, but will also involve some rather sophisticated
 stellar physics (see e.g. Hurley et al. 2001). Finally including an external tidal field
 is important for proper cluster modelling.

 We think that our way of modeling very large realistic star clusters is promising as a
 counterpart (presumably the only available) for the upcoming very large $N$-body simulations
 using recent generations of supercomputers (special and general purpose). Even with a few
 hundred Teraflops there will be no room for a large parameter survey in such models, which
 could be done with our stochastic Monte Carlo model.

\section*{Acknowledgments}
We would like to thank Douglas C. Heggie for stimulating discussions,
comments and suggestions.
This work was supported in part by the Polish National Committee for Science
Research under grant 2-P30-400-906. Financial Support by German Science
 Foundation under grant No. 436 POL 17/10/00, and
 436 POL 18/3/99 is gratefully acknowledged.


\begin{thebibliography}{}
\bibitem[]{} Aarseth S.J., Zare K., 1974, Celestial Mechanics, 10, 185
\bibitem[]{} Aarseth S.J., 1985, in Brackbill J.U.,
   Cohen B.I., eds, Multiple time scales, Academic Press, Orlando,
      p. 378
\bibitem[1994]{r3} Aarseth S.J., 1994,
in G. Contopoulos, N. K. Spyrou, L. Vlahos, eds,
 Galactic dynamics and N-body simulations,
 Lect. Notes Phys., Vol. 433, Springer, Berlin, Germany, p. 277
\bibitem[1999]{r5} Aarseth S.J., 1999a, PASP, 111, 1333
\bibitem[1999]{s5} Aarseth S.J., 1999b, CeMDA, 73, 127
\bibitem[2001]{a1} Aarseth S.J., 2003, in J. Makino \& P. Hut (eds),
Astrophysical Supercomputing using Particle
Simulations, Proc. IAU Symp. 208, ASP IAU Symp. Volumes, in press, astro-ph/0110148
\bibitem[]{} Aarseth S.J., Heggie D.C., 1998, MNRAS, 297, 794
\bibitem[1973]{r6} Ahmad A., Cohen L., 1973, J. Comp. Phys., 12, 389
\bibitem[Bacon et al.~1996]{1996MNRAS.281..830B} Bacon D., Sigurdsson S.,
Davies M.~B., 1996, MNRAS,  281, 830
\bibitem[2001]{s7} Baumgardt H., 2001, MNRAS 325, 1323
\bibitem[2002]{}Baumgardt H., Makino J., 2002, PASJ, subm.
\bibitem[2001]{s8} Baumgardt H., Hut P., Heggie D.C., 2002, MNRAS, subm., astro-ph/0206258
\bibitem[]{} Bettwieser E., Sugimoto D., 1984, MNRAS, 208, 493
\bibitem[]{} Cohn H., 1979, ApJ, 234, 1036
\bibitem[]{} Cohn H., 1980, ApJ, 242, 765
\bibitem[]{} Cohn H., Hut P., Wise M., 1989, ApJ, 342, 814
\bibitem[]{} Dorband E.N., Hemsendorf M., Merritt D., 2002, subm. JCP, astro-ph/0112092
\bibitem[]{} Drukier G.A., Cohn H.N., Lugger P.M., Yong H., 1999, ApJ, 518, 233
\bibitem[]{} Einsel C., Spurzem R., 1999, MN, 302, 81
\bibitem[]{} Fregeau J.M., G\"urkan M.A., Joshi K.J., Rasio F.A., 2003, 
submitted to ApJ, astro-ph/0301521
\bibitem[2000]{s16} Freitag M., PhD Thesis 2000, Univ. de Gen\`eve, Switzerland
\bibitem[2001]{s16} Freitag M., Benz W., 2001, A\&A, 375, 711
\bibitem[2002]{s18} Freitag M., Benz W., 2002, in prep.
\bibitem[]{} Giersz M., 1996, in Hut P., Makino J., eds,
 Proc. IAU Symp. 174, Dynamics of Star Clusters, Reidel, Dordrecht
\bibitem[]{} Gao B., Goodman J., Cohn H., Murphy B., 1991, ApJ, 370, 567
\bibitem[1998]{r17} Giersz M., 1998, MNRAS, 298, 1239
\bibitem[2001]{s17} Giersz M., 2001, MNRAS, 324, 218
\bibitem[]{} Giersz M., Heggie D.C., 1994a, MNRAS, 268, 257
\bibitem[]{} Giersz M., Heggie D.C., 1994b, MNRAS, 270, 298
\bibitem[]{} Giersz M., Heggie D.C., 1996, MNRAS, 279, 1037
\bibitem[]{} Giersz M., Spurzem R., 1994, MNRAS, 269, 241
\bibitem[]{} Giersz M., Spurzem R., 2000, MNRAS, 317, 581, Paper II
\bibitem[]{} Grillmair C.J., Forbes D.A., Brodie J.P., Elson R.A.W.,
 1999, AJ, 117, 167
\bibitem[]{} Hamada T., Fukushige T., Kawai A., Makino J., 2000, PASJ, 52, 943
\bibitem[]{} Heggie D.C., 1975, MNRAS, 173, 729
\bibitem[]{} Heggie D.C., 1984, MNRAS, 206, 179
\bibitem[]{} Heggie D.C., 2000, MNRAS, 318, L61
\bibitem[]{} Heggie D.C., Aarseth S.J., 1992, MNRAS, 257, 513
\bibitem[Heggie \& Rasio 1996]{1996MNRAS.282.1064H} Heggie D.~C., Rasio
F.~A., 1996, MNRAS,  282, 1064
\bibitem[]{} Hemsendorf M., Sigurdsson S., Spurzem R., 2002, ApJ, 581, 1256
\bibitem[]{} H\'enon M., 1971, Ap\&SS, 14, 151
\bibitem[]{} H\'enon M., 1975, in A. Hayli, ed, Proc. IAU Symp. 69,
   Dynamics of Stellar Systems, Reidel, Dordrecht, p. 133
\bibitem[Hills 1975]{1975AJ.....80.1075H} Hills J.~G., 1975a, AJ,  80, 1075
\bibitem[Hills 1975]{1975AJ.....80..809H} Hills J.~G., 1975b, AJ,  80, 809
\bibitem[]{} Huang X., McMillan S.L.W., 2001, in S. Deiters, B. Fuchs, A. Just, R. Spurzem,
 R. Wielen, eds, Dynamics of Star Clusters and the Milky Way, ASP Conf. Ser. 228, p. 449
\bibitem[]{} Hurley J.R., Tout C.A., Aarseth S.J., Pols O.R., 2001, MNRAS, 323, 630
\bibitem[Hut \& Bahcall 1983]{1983ApJ...268..319H} Hut P., Bahcall J.~N.,
1983, ApJ,  268, 319
\bibitem[]{} Hut P., Inagaki S., 1985, ApJ, 298, 502
\bibitem[]{} Hut P., Makino J., 1999, Science, 283, 501
\bibitem[]{} Hut P., Paczy\'nski B., 1984, ApJ, 284, 675
\bibitem[]{}  Hut P., Shara M.M., Aarseth S.J., Klessen R.S., Lombardi J.C. Jr.,
Makino J., McMillan S., Pols O.R., Teuben P.J., Webbink R.F., 2002, NewA (subm.),
astro-ph/0207318
\bibitem[]{} Ibata R.A., Richer H.M., Fahlman G.G., Bolte M., Bond H.E.,
 Hesser J.E., Pryor C., Stetson P., 1999, ApJS, 120, 265
\bibitem[2000]{s32} Joshi K.J., Rasio F.A., Portegies Zwart S.F., 2000, ApJ, 540, 969
\bibitem[2001]{s32} Joshi K.J., Nave C.P., Rasio F.A., 2001, ApJ, 550, 691
\bibitem[Kim et al.~2002]{2002MNRAS.334..310K} Kim E., Einsel C., Lee
H.~M., Spurzem R., Lee M.~G., 2002, MNRAS,  334, 310
\bibitem[Kroupa 1995]{1995MNRAS.277.1491K} Kroupa P., 1995, MNRAS,  277,
1491
\bibitem[]{} Kuberka T., Kugel A., M"anner R., Singpiel H., Spurzem R., Klessen
 R., 1999, in ``Procs. of The International Conference on Parallel and
 Distributed Processing Techniques and Applications (PDPTA 99),
 Las Vegas, USA
\bibitem[]{} Kustaanheimo P., Stiefel E.J., 1965, J. Reine Angew. Math., 218, 204
\bibitem[]{} Lippert T., Glaessner U., Hoeber H., Ritzenh"ofer G., Schilling K.,
 Seyfried A., 1996, Int. J. Mod. Phys., 7, 485
\bibitem[]{} Lippert T., Seyfried A., Bode A., Schilling K., 1998, IEEE Transact. Par.
 Distr. Syst., 9, 1
\bibitem[]{} Louis P.D, Spurzem R., 1991, MNRAS, 251, 408
\bibitem[]{} Lupton R.H., Gunn J.E., Griffin R.F., 1987, AJ, 93, 1114
\bibitem[]{} Lynden-Bell D., Eggleton P.P., 1980, MNRAS, 191, 483
\bibitem[]{} Makino J., 1996, ApJ, 471, 796
\bibitem[]{} Makino, J., 2002, NewA, 7, 373
\bibitem[]{} Makino J., Aarseth S.J., 1992, PASJ, 44, 141
\bibitem[]{} Makino, J., Hut P., (eds.), 2003: Astrophysical Supercomputing using Particle
Simulations, Proc. IAU Symp. 208, ASP IAU Symp. Volumes, in press
\bibitem[]{} Makino, J., Taiji, M. 1998. Scientific simulations with
special purpose computers. Chichester: Wiley
\bibitem[]{} Makino J., Taiji M., Ebisuzaki T., Sugimoto D., 1997, ApJ, 480, 432
\bibitem[]{} Mardling R.A, Aarseth S.J., 2001, MNRAS, 3211, 398
\bibitem[]{} McMillan S.L.W., Hut P., Makino J., 1991, ApJ, 372, 111
\bibitem[]{} McMillan S.L.W., Hut P., 1996, ApJ, 467, 348
\bibitem[]{} Mikkola S., 1983, MNRAS, 203, 1107
\bibitem[]{} Mikkola S., 1984a, MNRAS, 207, 115
\bibitem[]{} Mikkola S., 1984b, MNRAS, 208, 75
\bibitem[]{} Mikkola S., Aarseth S.J. 1996, CeMDA 64, 197
\bibitem[]{} Mikkola S., Aarseth S.J. 1998, NewA 3, 309
\bibitem[]{} Milosavljevic M., Merritt D., 2001, ApJ, 563, 34
\bibitem[]{} Murphy B.W., Cohn H., Hut P., 1990, MNRAS, 245, 335
\bibitem[]{} Murphy B.W., Cohn H.N., Durisen R.H., 1991, ApJ, 370, 60
\bibitem[]{} Piotto G., Zoccali M., 1999, AA, 345, 485
\bibitem[]{} Piotto G., Zoccali M., King I.R., Djorgovski S.G., Sosin C.,
 Dorman B., Rich R.M., Meylan G., 1999, AJ, 117, 264
\bibitem[]{} Portegies Zwart S.F., et al., 1998, AA, 337, 363
\bibitem[Rasio \& Heggie 1995]{1995ApJ...445L.133R} Rasio F.~A., Heggie
D.~C., 1995, ApJL,  445, L133
\bibitem[Rubenstein \& Bailyn 1997]{1997ApJ...474..701R} Rubenstein E.~P.,
Bailyn C.~D., 1997, ApJ,  474, 701
\bibitem[]{} Shara M.M., Fall S.M., Rich R.M., Zurek D., 1998, ApJ, 508, 570
\bibitem[Sigurdsson \& Phinney 1993]{1993ApJ...415..631S} Sigurdsson S.,
Phinney E.~S., 1993, ApJ,  415, 631
\bibitem[]{} Sills A., Deiters S., Eggleton P., Freitag M., Giersz M., Heggie D.C.,
Hurley J., Hut P., Ivanova N., Klessen R., Kroupa P., Lombardi J. C., McMillan S.L.W.,
Portegies Zwart S.F., Zinnecker H., 2003, astro-ph/0301478
\bibitem[]{} Spitzer L., 1975, in A. Hayli (ed.)
     Dynamcis of Stellar Systems, Proc. IAU Symp. No. 69, p. 3.
     Reidel, Dordrecht
\bibitem[]{} Spitzer L., 1987, Dynamical Evolution of Globular
  Clusters, Princeton Univ. Press, Princeton
\bibitem[]{} Spurzem R., 1994, in Pfenniger D., Gurzadyan V.G., eds,
 Ergodic Concepts in Stellar Dynamics, Springer-Vlg.,
 Berlin, Heidelberg, p. 170
\bibitem[]{} Spurzem R., 1996, in Hut P., Makino J., eds, Dynamics
 of Star Clusters, Proc. IAU Symp. No. 174, p. 111
\bibitem[1999]{r55} Spurzem R., 1999, in Riffert H., Werner K. (eds),
Computational Astrophysics, The Journal of Computational and Applied Mathematics
   (JCAM) 109, Elsevier Press, Amsterdam, p.\ 407
\bibitem[]{} Spurzem R., Aarseth S.J., 1996, MNRAS, 282, 19
\bibitem[1996]{r57} Spurzem R., Giersz M., 1996, MNRAS, 283, 805, Paper I
\bibitem[1999]{s57} Spurzem R., Kugel A., 1999, in Procs. of NIC Workshop
 on Molecular Dynamics on Parallel Computers, World Scientific, Singapore.
\bibitem[]{} Spurzem R., Takahashi K., 1995, MNRAS, 272, 772
\bibitem[2002]{s56} Spurzem R., Makino J., Fukushige T., Lienhart G., Kugel A.,
 M\"anner R., Wetzstein M., Burkert A., Naab T., 2002, in
Intl. Symp. on Comp. Science \& Engineering in the Next Decade,
Fifth JSPS/CSE Symposium, Tokyo 2002, in press
\bibitem[]{} \Stodolkiewicz, J.S. 1982, Acta Astron. 32, 63
\bibitem[]{} \Stodolkiewicz J.S., 1985, in Goodman J., Hut P., eds, Proc.
   IAU Symp. 113, Dynamics of Star Clusters, Reidel, Dordrecht, p. 361
\bibitem[]{} \Stodolkiewicz, J.S. 1986, Acta Astron. 36, 19
\bibitem[]{} Sugimoto D., Chikada Y., Makino J., Ito T.,
    Ebisuzaki T., Umemura M., 1990, Nature, 345, 33
\bibitem[Takahashi 1995]{1995PASJ...47..561T} Takahashi K., 1995, PASJ,
47, 561
\bibitem[Takahashi 1996]{1996PASJ...48..691T} Takahashi K., 1996, PASJ,
48, 691
\bibitem[Takahashi 1997]{1997PASJ...49..547T} Takahashi K., 1997, PASJ,
49, 547
\bibitem[Takahashi \& Portegies Zwart 1998]{1998ApJ...503L..49T} Takahashi
K., Portegies Zwart S.~F., 1998, ApJL,  503, L49
\bibitem[Takahashi \& Portegies Zwart 2000]{2000ApJ...535..759T} Takahashi
K., Portegies Zwart S.~F., 2000, ApJ,  535, 759
\bibitem[2000]{s61} Watters W.A., Joshi K.J., Rasio F.A., 2000, ApJ, 539, 331
 
 \end{thebibliography}
 \end{document}